\documentstyle[12pt,psfig,a4]{article}
\begin{document}
\def\doublespaced{\baselineskip=\normalbaselineskip\multiply\baselineskip 
  by 150\divide\baselineskip by 100}
\begin{titlepage}
\vspace{0.5cm}
\begin{flushright}
\large
hep-ph/9906215 \\
June 1999 \hfill MSUHEP-90523 
\end{flushright}
\vspace{0.5cm}
\begin{center}
\large
{New Physics in the Third Family and its Effect on Low Energy Data}
\end{center}  
\begin{center}
{\bf Ehab Malkawi$^{a,}$\footnote{e-mail:malkawie@just.edu.jo} 
and C.--P. Yuan$^b$}
\end{center}
\begin{center}
{$^a$Department of Physics,
Jordan University of Science \& Technology\\
 Irbid 22110, Jordan}
\end{center}
\begin{center}
{$^b$Department of Physics and Astronomy,
Michigan State University \\
East Lansing, MI 48824, USA}
\end{center} 
\vspace{0.4cm}
\raggedbottom 
\relax
\begin{abstract}
\noindent
We investigate, in detail, a model in which the third family fermions are 
subjected to an SU(2) dynamics different from the first two families. 
Constrained by the precision $Z$-pole data, the heavy gauge boson mass 
is bounded from below to be about $1.7$ TeV at the $2\sigma$ level. 
The flavor-changing neutral current (FCNC) in the lepton sector can be 
significant in $\tau\leftrightarrow e$ and $\tau \leftrightarrow \mu$ 
transitions. In the latter case, the ratio 
${\rm{Br}}(\tau\rightarrow \mu \overline{\nu_\mu} \nu_\tau)/
{\rm{Br}}(\tau\rightarrow e \overline{\nu_e} \nu_\tau)$ and 
${\rm{Br}}(\tau\rightarrow \mu \mu \mu)$
can constrain the model better than LEP/SLC data in some region of the 
parameter space. Furthermore, FCNCs are unavoidable in the quark sector. 
Significant effects to the $B^0$-$\overline{B^0}$ mixing and the rare 
decays of the $K$ and $B$ mesons, such as 
$K^\pm \to \pi^\pm \nu {\overline \nu}$, 
$b \to s \nu {\overline \nu}$, $B_s \to \tau^+\tau^-$, $\mu^+\mu^-$ and
$B_{s,d} \to \mu^\pm \tau^\mp$,  are expected.
\end{abstract} 

\vspace{0.5cm} \noindent PACS numbers:12.15.Ji, 12.60.-i, 12.60.Cn, 13.20.-
v, 13.35.-r \vspace{1.0cm} \end{titlepage} \newpage 


\setcounter{page}{1}
\pagenumbering{arabic}
\pagestyle{plain}
\def\sm{\mbox{$SU(3)_C\times SU(2)_L \times U(1)_Y$}\,}
\def\suu{$SU(2)_L \times U(1)_Y$\,}
\def\su{$SU(2)_l \times SU(2)_h\times U(1)_Y$\,}
\def\slash{\not{}{\mskip-3.mu}}
\def\ra{\rightarrow}
\def\lra{\leftrightarrow}
\def\bea{\begin{eqnarray}}
\def\ena{\end{eqnarray}}
\def\beq{\begin{equation}}
\def\enq{\end{equation}}
\def\cs{\cos{\theta}}
\def\sn{\sin{\theta}}
\def\tw{\tan\theta}
\def\css{\cos^2{\theta}\,}
\def\sns{\sin^2{\theta}\,}
\def\tns{\tan^2{\theta}\,}
\def\eg{{\it e.g.},\,\,}
\def\ie{{\it i.e.},\,\,}
\def\etc{{\it etc}}
\def\MWs{M^2_W}
\def\MZs{M^2_Z}
\def\MHs{m^2_H}
\def\mf{m_f}
\def\mb{m_b}
\def\mt{m_t}
\def\snshel{\sin^2{\hat{\theta}}\,}
\def\csshel{\cos^2{\hat{\theta}}\,}
\def\snsms{\overline{s}^2_{\theta}\,}
\def\cssms{\overline{c}^2_{\theta}\,}
\def\kln{\kappa_{L}^{NC}}
\def\krn{\kappa_{R}^{NC}}
\def\klc{\kappa_{L}^{CC}}
\def\krc{\kappa_{R}^{CC}}
\def\ttz{{\mbox {\,$t$-${t}$-$Z$}\,}}
\def\bbz{{\mbox {\,$b$-${b}$-$Z$}\,}}
\def\tta{{\mbox {\,$t$-${t}$-$A$}\,}}
\def\bba{{\mbox {\,$b$-${b}$-$A$}\,}}
\def\tbw{{\mbox {\,$t$-${b}$-$W$}\,}}
\def\tltlz{{\mbox {\,$t_L$-$\overline{t_L}$-$Z$}\,}}
\def\blblz{{\mbox {\,$b_L$-$\overline{b_L}$-$Z$}\,}}
\def\brbrz{{\mbox {\,$b_R$-$\overline{b_R}$-$Z$}\,}}
\def\tlblw{{\mbox {\,$t_L$-$\overline{b_L}$-$W$}\,}}
\def\ppbar{ \bar{{\rm p}} {\rm p} }
\def\pp{ {\rm p} {\rm p} }
\def\ipb{ {\rm pb}^{-1} }
\def\ifb{ {\rm fb}^{-1} }
\def\stds{\strut\displaystyle}
\def\SST{\scriptscriptstyle}
\def\TT{\textstyle}
\def\D0{D\O~}
\def\sqrts{\sqrt{s}}
\def\lsim{~{\rlap{\lower 3.5pt\hbox{$\mathchar\sim$}}\raise 1pt\hbox{$<$}}\,}
\def\gsim{~{\rlap{\lower 3.5pt\hbox{$\mathchar\sim$}}\raise 1pt\hbox{$>$}}\,}

\section{Introduction}

\indent

The search for physics beyond the Standard Model (SM) is an ongoing
endeavor. Usually, a search for new physics implies investigating higher
and higher energy regime where new physics effects are expected to
appear. Nevertheless, it remains a necessary and useful approach to study the 
low-energy regime where interesting phenomena may be expected in a particular 
model. The work presented here is an example where new
physics diminishes in some of the very low-energy processes and flourishes 
in the others.

The flavor physics of the third generation is particularly mysterious for 
the smallness of the mixing angles and the huge hierarchy
in masses. Furthermore, the heavy top quark mass can be an indication for 
a new dynamics in the third fermion generation 
different from the first two generations. 
It is interesting to investigate the idea of treating the third generation 
differently from the first two generations in the context of strong or 
electroweak interactions.
Fortunately, the already available low energy data can largly constrain such 
a picture. In this regard, several studies have been pursued in the 
literature. In the context of the Quantum Chromodynamics (QCD) interaction, 
we refer the reader to Refs.~\cite{hill1, hill2}. 
In the context of the electroweak 
interaction, several published works also exist. As an example, 
in the context of Tecnicolor theories, we refer the reader to 
Ref.~\cite{chiv}. 
The idea that the third generation
carries a seperate $SU(2)$ was proposed in Refs.~\cite{lima, ehab5, muller}. 
The two models in Refs.~\cite{ehab5, muller}
differ in the assignment of the quantum numbers to the Higgs sector which 
leads to different phenomenological implications. 
Constraints from low energy data on such models have been 
discussed in Refs.~\cite{ehab5, muller, lee, andrea}. 

In Ref.~\cite{ehab5}, we proposed a model in which  
the third generation feels a different
gauge dynamics (with a new $SU(2)$ gauged symmetry)
from the usual weak interaction proposed in the SM. 
(No modification to the QCD interaction was considered, because that 
case has been discussed elsewhere \cite{hill1, hill2}.)
Consequently, a new spectrum of gauge bosons emerges in the model. We, then,
used the available CERN Large Electron Positron (LEP) and 
SLAC Stanford Linear Collider (SLC) data to constrain the parameters of the
model. We found the model to be consistent with data (at the
3$\sigma $ level) as long as the heavy gauge boson mass is larger than 1.3
TeV.
A similar conclusion was also found in 
Refs.~\cite{muller,lee,andrea}.\footnote{
Though, the assignment of the fermion quantum numbers may not be
identical.}

In this current work, we first update the previous analysis on constraining
the parameter space of the proposed model
using the most recent LEP and SLC data \cite{lep}, then
discuss the zero-momentum transfer physics in the low-energy 
regime where interesting effects
may be expected in both lepton and quark sectors. 
We find that flavor-changing neutral currents (FCNCs) may exist 
in the lepton sector and are unavoidable in the quark sector.
As a consequence, neutrinos can mix via gauge interaction despite of their 
zero mass. Furthermore, deviations from the SM predictions 
are expected for some particular low-energy processes. For example, the 
decay process $\tau \longrightarrow \mu 
\overline{\nu _\mu }\nu _\tau $ can impose a 
stronger constraint than the $Z$-pole 
data for some particular parameter space. 
Similarly, the $B^0$-$\overline{B^0}$ mixing and the rare decay rates of 
the $K$ and $B$ mesons, such as
$K^\pm \to \pi^\pm \nu {\overline \nu}$ and 
$B_{s} \to \tau^+ \tau^-, \mu^+ \mu^-$
are expected to exceed 
the SM prediction for some region of the parameter space.
Non-SM decay modes, such as $B_{s,d} \to \mu^\mp \tau^\pm$, can also
occur.

The rest of this paper is organized as follows.
In Sec. 2, we briefly review the model. In Sec. 3, we discuss the 
constraints on the model from the $Z$-pole data at LEP and SLC.
After a general discussion on the possible new effects on low energy data 
in Sec. 4, we discuss 
all possible new physics effects, including all FCNC processes as predicted 
by this model, in Secs. 5 and 6. We 
summarize our conclusions in Sec. 7.

\section{The Model}

\indent

For the detailed structure of the model, we refer the reader to 
Ref.~\cite{ehab5}. 
In this section we only outline the main features of the proposed
model. The model is based on the gauge symmetry G=~$SU(2)_l\times
SU(2)_h\times U(1)_Y$\thinspace. The third generation of fermions (top
quark, $t$, bottom quark, $b$, tau lepton, $\tau $, and
 its neutrino, $\nu
_\tau $) experiences a new gauge interaction, instead of the usual weak
interaction advocated by the SM. On the contrary, the first and second
generations only feel the weak interaction supposedly equivalent to the SM
case. The new gauge dynamics is attributed to the $SU(2)_h$ symmetry
under which the left-handed fermions of the third generation transform in
the fundamental representation (doublets), while they remain to be singlets
under the $SU(2)_l$ symmetry. On the other hand, the left-handed fermions of
the first and second generation transform as doublets under the $SU(2)_l$
group and singlets under the $SU(2)_h$ group. The $U(1)_Y$ group is the SM
hypercharge group. The right-handed fermions only transform under the 
$U(1)_Y $ group as assigned by the SM. Finally the QCD interactions and the
color symmetry $SU(3)_C$ are the same as that in the SM.

The symmetry breaking of the Lie group G into the electromagnetic group 
$U(1)_{{\rm {em}}}$ is a two-stage mechanism. 
First, $SU(2)_l \times
SU(2)_h\times U(1)_Y$\, breaks down into $SU(2)_L\times U(1)_Y$ at some
large energy scale. The second stage is that
 $SU(2)_L\times U(1)_Y$ breaks
down into $U(1)_{{\rm {em}}}$ at an energy scale about
the same as the SM electroweak
symmetry-breaking scale. The spontaneous symmetry-breaking of the group 
$SU(2)_l \times SU(2)_h\times U(1)_Y$\, is accomplished by introducing two
scalar matrix fields $\Sigma$ and $\Phi$    
which transform as 
\begin{eqnarray*}
\Sigma\sim (2,2)_0 \,\, ,\hspace{1cm} \Phi \sim (2,1)_1\,\,\, ,
\end{eqnarray*}
i.e., the $\Sigma$ field transforms as a doublet under both 
$SU(2)_l$ and $SU(2)_h$ and as a singlet under $U(1)_Y$. On the other hand, 
the $\Phi$ field transforms as a doublet under $SU(2)_l$, as a singlet under 
$SU(2)_h$, and its hypercharge quantum number $Y$ is 1. 
Thus, the scalar 
doublet $\Phi$ carries equivalent quantum numbers as the SM Higgs doublet. 

As a realization of the symmetry,
the $\Sigma$ and $\Phi$ fields transform as 
\begin{eqnarray*}
\Sigma \rightarrow g_1\Sigma g_2^{\dagger }\,\,,\hspace{1cm}\Phi \rightarrow
g_1g_Y\Phi \,\,,
\end{eqnarray*}
where $g_1\in SU(2)_l$, $g_2\in SU(2)_h$, and $g_Y\in U(1)_Y$. 
For completeness, we briefly discuss 
the structure of the boson and lepton sectors as follows.

\subsection{The Boson Sector}

\indent

The covariant derivatives of the scalar fields
are defined as 
\begin{eqnarray}
D^\mu \Sigma =\partial^\mu \Sigma +ig_lW_l^\mu \Sigma -ig_h\Sigma W_h^\mu\,,
\nonumber \\
D^\mu \Phi =\partial^\mu \Phi +ig_lW_l^\mu \Phi +\frac{i}{2}g^{\prime }B^\mu
\Phi \,,
\end{eqnarray}
where $W_{l,h}\equiv W_{l,h}^a\tau ^a/2$ and where $W_{l,h}^a$ are
the gauge boson fields of the $SU(2)_{l,h}$ groups,
respectively.
($\tau^a$'s are
the Pauli matrices, and ${\rm {Tr}}(\tau^a \tau^b)=2\delta_{ab}$.)

With these definitions, the gauge invariant Lagrangian 
of the boson sector is 
\begin{eqnarray}
{\cal {L}}_B &=& \frac{1}{2} D_\mu \Phi^{\dagger} D^\mu \Phi +
 \frac{1}{4}{\rm {Tr}}(D_\mu \Sigma^{\dagger }D^\mu \Sigma )+
{\rm {V}}(\Phi ,\Sigma )  \nonumber\\
&&-\frac{1}{4} {W_l^a}_\mu {W_l^a}^\mu -\frac{1}{4}{W_h^a}_\mu {W_h^a}^\mu -
\frac{1}{4}B_\mu B^\mu \,\,,
\end{eqnarray}
where ${\rm {V}}(\Phi ,\Sigma )$ is the scalar potential. We assume that the
first stage of symmetry breaking is accomplished through the $\Sigma $
field by acquiring a vacuum expectation value (VEV) $\,u$, i.e.,
$
\left\langle \Sigma \right\rangle =\pmatrix{u & 0 \cr 0 & u \cr}\,.
$
The second stage is 
through the scalar $\Phi $ field by
acquiring a vacuum expectation value $v$, so 
$
\left\langle \Phi \right\rangle=\pmatrix{0 \cr v\cr}\, ,
$
where $v$ is at the same order as the SM symmetry-breaking scale. Because 
of this
pattern of symmetry breaking, the gauge couplings are related to the 
$U(1)_{{\rm {em}}}$ gauge coupling $e$ by the relation 
\begin{equation}
\frac 1{e^2}=\frac 1{g_l^2}+\frac 1{g_h^2}+\frac 1{{g^{\prime }}^2}\,\,.
\end{equation}
We then define 
\begin{equation}
g_l=\frac e{\sin \theta \cos \phi }\,,\hspace{1cm} 
g_h=\frac e{\sin \theta \sin \phi }\,,\hspace{1cm}
g^{\prime }=\frac e{\cos \theta }\,,
\end{equation}
where $\theta$ plays the role of the usual weak mixing 
angle and $\phi $ is a new
parameter of the model. The scalar fields, except ${\rm {Re}}(\phi ^0)$
from the $\Phi $ doublet and $\sigma $ from the 
$\Sigma(\equiv \sigma +i\pi^a\tau^a)$ matrix field,
become the longitudinal components of the physical gauge bosons. The
surviving ${\rm {Re}}(\phi ^0)$ field behaves similar to the SM Higgs boson
except that it does not have the usual Yukawa couplings to the third
generation. 

To derive the mass eigenstates and physical masses of the gauge bosons, we
need to diagonalize their mass matrices. 
For $g_h>g_l$ (equivalently $\tan \phi <1$), we require 
$g_h^2\leq 4\pi $ (which implies $\sin ^2\phi \geq g^2/(4\pi )\sim 1/30$) so
that the perturbation theory is valid. Similarly, for $g_h<g_l$, we require 
$\sin ^2\phi \leq 0.96$. For simplicity, we focus on the region where 
$x(\equiv u^2/v^2)$ is much larger than 1, and
ignore the corrections which are suppressed by higher powers of $1/x$. 
To the order $1/x$, the light gauge boson masses are found to be \cite{ehab5} 
\begin{equation}
M^2_{W^\pm }=M_0^2(1-\frac{\sin ^4\phi }x)\,,
\end{equation}
\begin{equation}
M_Z^2=\frac{M_0^2}{\cos ^2{\theta }\,}(1-\frac{\sin ^4\phi }x)\,,
\label{chapter5_mz}
\end{equation}
where $M_0 \equiv ev/2\sin \theta $.
While for the heavy gauge bosons, one finds 
\begin{equation}
M^2_{{W^\prime}^\pm} = M^2_{Z^\prime }=M_0^2\left( \frac x{\sin^2\phi
\cos^2\phi }+\frac{\sin^2\phi }{\cos^2\phi }\right) \,.
\end{equation}
It is interesting to notice that up to this order
the heavy gauge bosons are degenerate in mass.
This is because the heavy gauge bosons
do not mix with the hypercharge gauge boson field, $B_\mu $.

\subsection{The Fermion Sector}

\indent

As discussed before, the third generation interacts
with the $SU(2)_h$ gauge bosons, and the first and second generations 
interact with the $SU(2)_l$ gauge bosons. Explicitly, under the 
$SU(2)_l\times SU(2)_h\times U(1)_Y$ symmetry,
the quantum numbers of the fermions are assigned as follows.
For the first and second generation fermions, we assign \\
left-handed quarks : $(2,1)_{1/3}$ \thinspace , \hspace{1cm} 
left-handed leptons : $(2,1)_{-1}$ \thinspace . \\
For the third generation, we have \\
left-handed quarks : $(1,2)_{1/3}$ \thinspace , \hspace{1cm}
left-handed leptons : $(1,2)_{-1}$ \thinspace . \\
For all the right-handed fermions, we assign \\
right-handed quarks and leptons : $(1,1)_{2Q}$ \thinspace , \\
where $Q$ is the electric charge of the fermions. 
Because of this assignment, the model is anomaly free,
and the cancellation of
anomalies is satisfied family by family.

In terms of the mass eigenstates of the gauge bosons $W^{\pm}$, $Z$, 
${W^\prime}^\pm$, and $Z^{\prime}$, the fermionic interaction Lagrangian is
\begin{eqnarray}
&&{\cal L}_f^{int}= \frac{e}{\sin\theta}\overline{\Psi_L}\gamma^{\mu} 
\left [ T_l^\pm + T_h^\pm +\frac{\sin^2\phi}{x} \left( 
T_h^\pm \cos^2\phi-T_l^\pm\sin^2\phi\right ) \right ] \Psi_L W_\mu^\pm +  
\nonumber \\
&&\frac{e}{\sin\theta\cos \theta}\overline{\Psi_L}\gamma^{\mu} 
\left [T_l^3+ T_h^3 -Q \sin^2\theta +\frac{\sin^2\phi}{x} \left(
\cos^2\phi T_h^\pm-\sin^2\phi T_l^\pm\right) \right ] \Psi_L Z_\mu  
\nonumber \\
&&+\frac{e}{\sin\theta}\overline{\Psi_L}\gamma^{\mu} 
\left [- \frac{\sin\phi}{\cos \phi}T_l^\pm + 
\frac{\cos\phi}{\sin \phi}T_h^\pm -
\frac{\sin^3\phi\cos\phi}{x\cos^2\theta} 
\left(T_h^\pm + T_l^\pm\right) \right ]
\Psi_L {W_\mu^\prime}^\pm +  \nonumber \\
&&\frac{e}{\sin\theta}\overline{\Psi_L}\gamma^{\mu} 
\left [- \frac{\sin\phi}{\cos \phi}T_l^3 + \frac{\cos\phi}{\sin \phi}T_h^3 -
\frac{\sin^3\phi\cos\phi}{x\cos^2\theta} 
\left( T_h^3 + T_l^3-Q\sin^2\theta \right)\right ] \Psi_L Z_\mu^{\prime}  
\nonumber \\
&&+eQ\overline{f}^i\gamma^\mu f^i A_\mu -\frac{eQ\sin^2\theta}
{\sin\theta\cos\theta}\overline{f_R}^i\gamma^\mu f_R^i\left (Z_\mu - 
\frac{\sin^3\phi \cos\phi}{x\cos\theta}Z_\mu^{\prime} \right )\, .
\end{eqnarray}

The first and second
generations acquire their masses through the Yukawa interactions to the 
$\Phi $ doublet field. The fermions Yukawa Lagrangian is 
\begin{eqnarray}
{\cal L}_{\mbox{\rm{Yukawa}}} &=&\overline{\Psi _L}^1\Phi \left[
g_{11}^ee_R+g_{12}^e\mu _R+g_{13}^e\tau _R\right] + \nonumber \\
&&\ \overline{\Psi _L}^2\Phi \left[ g_{21}^ee_R+g_{22}^e\mu _R+g_{23}^e\tau
_R\right] +h.c.  \label{chap5_mass_1}
\end{eqnarray}
For the third generation one cannot generate their masses through the
usual Yukawa terms (dimension-four operators), as it is not allowed by gauge
invariance. This can be an indication that the mass generation of the third 
family is due to a different mechanism than the first two generations. 
One way to realize this is to assume that our proposed symmetry can be 
embedded in a larger symmetry at a much higher 
energy scale. The breaking of the large symmetry is responsible for the 
generation of the third family masses as it is also responsible for the new 
non-universal gauge dynamics. At the low energy scale this can be effectively 
written in terms of higher dimension operators. 
For example, the mass of the 
the $\tau$ lepton can be generated through the following 
dimension-five operators:
\begin{equation}
\frac 1\Lambda \overline{\Psi _L}^3\Sigma ^{\dagger }\Phi \left[
g_{31}e_R+g_{32}\mu _R+g_{33}\tau _R\right] +h.c.,  \label{chap5_mass_2}
\end{equation}
where $\Psi _L^3={\pmatrix{\nu _{\tau L}  \cr \tau_L \cr}}$, and $\Lambda $
characterizes some large mass scale associated with the strong flavor
interaction. It is reasonable to assume $\Lambda \sim u\gg v$, so that
the mass of $\tau$ is about equal to $g_{33}\, v$. 
Thus, although the masses of
the first and second generations are generated through the Yukawa
interactions as in the SM, the mass spectrum of third generation must be
generated by a different mechanism. 
This conclusion may be attributed
to strong flavor dynamics which could be evident at
high energies, where the interactions become strong. 
Another scenario \cite{lima} for generating the third
family masses in this model is to introduce 
an additional scalar doublet which only couples
to the third generation through the usual Yukawa interactions.
In general, this scenario will introduce extra 
interaction terms to the gauge dynamics and will
modify the conclusions presented in this paper.

Given the fermion mass matrices, one can derive their physical
masses by diagonalizing the mass matrices using bilinear unitary
transformations. For example, for the lepton sector, the lepton 
mass matrix $M_e$ can be read out from the Lagrangian written 
above in Eqs.~(\ref{chap5_mass_1}) and (\ref{chap5_mass_2}). We introduce the 
unitary matrices $L_e$ and $R_e$ with the transformations: 
\begin{equation}
e_L^i\rightarrow L_e^{ij}e_L^j\,,\hspace{1cm}e_R^i\rightarrow
R_e^{ij}e_R^j\,.
\end{equation}
Hence, the physical mass matrix is given by 
\begin{equation}
M_e^{{\rm {diag.}}}=L_e^{\dagger }M_eR_e\,.
\end{equation}
Because the third family interacts differently from the first and second
generation, we expect in general 
flavor-changing neutral currents to
occur at tree level.

In terms of the fermion mass eigenstates, the left-handed
neutral-current interactions are 
\begin{eqnarray*}
\frac e{2\sin \theta \cos \theta }\left( 
\begin{array}{ccc}
\overline{e_L} & \overline{\mu _L} & \overline{\tau _L}
\end{array}
\right) \gamma ^\mu \left[ -1+2\sin ^2\theta +\frac{\sin ^4\phi }x-
\frac{\sin ^2\phi }xL_e^{\dagger }GL_e\right] \left( 
\begin{array}{c}
e_L \\ 
\mu _L \\ 
\tau _L
\end{array}
\right) Z_\mu \, ,
\end{eqnarray*}
\begin{eqnarray*}
\frac e{2\sin \theta \cos \theta }\left( 
\begin{array}{ccc}
\overline{{\nu _e}_L} & \overline{{\nu _\mu }_L} & \overline{{\nu _\tau }_L}
\end{array}
\right) \gamma ^\mu \left[ 1-\frac{\sin ^4\phi }x+\frac{\sin ^2\phi }
xL_e^{\dagger }GL_e\right] \left( 
\begin{array}{c}
{\nu _e}_L \\ 
{\nu _\mu }_L \\ 
{\nu _\tau }_L
\end{array}
\right) Z_\mu \, ,
\end{eqnarray*}
\begin{eqnarray*}
\frac e{2\sin \theta}\left( 
\begin{array}{ccc}
\overline{e_L} & \overline{\mu _L} & \overline{\tau _L}
\end{array}
\right) \gamma ^\mu \left[ \frac{\sin\phi}{\cos\phi}+ 
\frac{\sin^3\phi \cos\phi}{x\cos^2\theta}(1-2\sin ^2\theta)-
\frac{L_e^{\dagger}GL_e}{\sin\phi\cos\phi} \right] \left( 
\begin{array}{c}
e_L \\ 
\mu _L \\ 
\tau _L
\end{array}
\right) Z^\prime_\mu \,,
\end{eqnarray*}
\begin{equation}
\frac e{2\sin \theta}\left( 
\begin{array}{ccc}
\overline{{\nu _e}_L} & \overline{{\nu _\mu }_L} & \overline{{\nu _\tau }_L}
\end{array}
\right) \gamma ^\mu \left[-\frac{\sin\phi}{\cos\phi}- 
\frac{\sin^3\phi \cos\phi}{x\cos^2\theta} +
\frac{L_e^{\dagger}GL_e}{\sin\phi\cos\phi} \right] \left( 
\begin{array}{c}
{\nu _e}_L \\ 
{\nu _\mu }_L \\ 
{\nu _\tau }_L
\end{array}
\right) Z^\prime_\mu \, ,
\end{equation}
where
\begin{equation}
G=\pmatrix{0 & 0 & 0 \cr 0 & 0 & 0 \cr 0 & 0 & 1 \cr}\,.  \label{G}
\end{equation}

The left-handed charged-current interactions are
 \begin{eqnarray*}
\frac e{\sqrt{2}\sin \theta }\left( 
\begin{array}{ccc}
\overline{e_L} & \overline{\mu _L} & \overline{\tau _L}
\end{array}
\right) \gamma ^\mu \left[ 1-\frac{\sin ^4\phi }x+\frac{\sin ^2\phi }
xL_e^{\dagger }GL_e\right] \left( 
\begin{array}{c}
{\nu _e}_L \\ 
{\nu _\mu }_L \\ 
{\nu _\tau }_L
\end{array}
\right) W_\mu ^{-}+\, h.c.,
\end{eqnarray*}
\begin{equation}
\frac e{\sqrt{2}\sin \theta }\left( 
\begin{array}{ccc}
\overline{e_L} & \overline{\mu _L} & \overline{\tau _L}
\end{array}
\right) \gamma ^\mu \left[- \frac{\sin\phi}{\cos\phi}- 
\frac{\sin^3\phi \cos\phi}{x\cos^2\theta} +
\frac{L_e^{\dagger}GL_e}{\sin\phi\cos\phi} \right] \left( 
\begin{array}{c}
{\nu _e}_L \\ 
\nu {_\mu }_L \\ 
{\nu _\tau }_L
\end{array}
\right) {W_\mu ^\prime}^- \, +{h.c.} 
\end{equation}
Similarly, for the quark sector we introduce the unitary matrices $L_u$ and 
$L_d$. In terms of the mass eigenstates one finds the following interaction
terms: 
\begin{eqnarray*}
\frac e{2\sin \theta \cos \theta }\left( 
\begin{array}{ccc}
\overline{u_L} & \ \overline{c_L} & \overline{t_L}
\end{array}
\right) \gamma ^\mu \left[ 1-\frac 43\sin ^2\theta -\frac{\sin ^4\phi }x+
\frac{\sin ^2\phi }xL_u^{\dagger }GL_u\right] \left( 
\begin{array}{c}
u_L \\ 
c_L \\ 
t_L
\end{array}
\right) Z_\mu ,
\end{eqnarray*}
\begin{eqnarray*}
\frac e{2\sin \theta \cos \theta }\left( 
\begin{array}{ccc}
\overline{d_L} & \ \overline{s_L} & \overline{b_L}
\end{array}
\right) \gamma ^\mu \left[ -1+\frac 23\sin ^2\theta +\frac{\sin ^4\phi }x-
\frac{\sin ^2\phi }xL_d^{\dagger }GL_d\right] \left( 
\begin{array}{c}
d_L \\ 
s_L \\ 
b_L
\end{array}
\right) Z_\mu \, ,
\end{eqnarray*}
\begin{eqnarray*}
\frac e{2\sin \theta}\left( 
\begin{array}{ccc}
\overline{u_L} & \ \overline{c_L} & \overline{t_L}
\end{array}
\right) \gamma ^\mu \left[ -\frac{\sin\phi}{\cos\phi}- 
\frac{\sin^3\phi \cos\phi}{x\cos^2\theta}(1-\frac{4}{3}\sin^2\theta) +
\frac{L_u^{\dagger}GL_u}{\sin\phi\cos\phi}\right] \left( 
\begin{array}{c}
u_L \\ 
c_L \\ 
t_L
\end{array}
\right) Z^\prime_\mu \,,
\end{eqnarray*}
\begin{eqnarray*}
\frac e{2\sin \theta}\left( 
\begin{array}{ccc}
\overline{d_L} & \ \overline{s_L} & \overline{b_L}
\end{array}
\right) \gamma ^\mu \left[ \frac{\sin\phi}{\cos\phi}+
\frac{\sin^3\phi \cos\phi}{x\cos^2\theta}(1-\frac{2}{3}\sin^2\theta) -
\frac{L_d^{\dagger }GL_d}{\sin\phi\cos\phi}\right] \left( 
\begin{array}{c}
d_L \\ 
s_L \\ 
b_L
\end{array}
\right) Z^\prime_\mu \, ,
\end{eqnarray*}
\begin{eqnarray*}
\frac e{\sqrt{2}\sin \theta }\left( 
\begin{array}{ccc}
\overline{u_L} & \ \overline{c_L} & \overline{t_L}
\end{array}
\right) \gamma ^\mu \left[ (1-\frac{\sin ^4\phi }x)L_u^{\dagger }L_d+
\frac{\sin ^2\phi }xL_u^{\dagger }GL_d\right] \left( 
\begin{array}{c}
d_L \\ 
s_L \\ 
b_L
\end{array}
\right) W_\mu ^{+}\,+h.c.,
\end{eqnarray*}
\begin{equation}
\frac e{\sqrt{2}\sin \theta }\left( 
\begin{array}{ccc}
\overline{u_L} & \ \overline{c_L} & \overline{t_L}
\end{array}
\right) \gamma ^\mu \left[ \left(-\frac{\sin\phi}{\cos\phi}- 
\frac{\sin^3\phi \cos\phi}{x}\right)L_u^\dagger L_d +
\frac{L_u^{\dagger}GL_d}{\sin\phi\cos\phi} \right ]
\left( 
\begin{array}{c}
d_L \\ 
s_L \\ 
b_L
\end{array}
\right) {W_\mu ^\prime}^+ \,+\, h.c. \label{qu2}
\end{equation}
The right-handed fermion couplings to the neutral gauge bosons $Z$ and 
$Z^{\prime }$ are, respectively, given by
\begin{eqnarray*}
\frac e{2\sin \theta \cos \theta }\left( -2Q\sin ^2\theta \right) ,
\end{eqnarray*}
\begin{equation}
\frac e{2\sin \theta }\left( 2Q\sin ^2\theta \frac{\sin ^3\phi \cos \phi }
{x\cos ^2\theta }\right) .
\end{equation}
The fermion couplings to the photon are the usual electromagnetic couplings.
As shown above, it is evident that
if $g_h>g_{{\em l}}$, then the heavy gauge bosons would
couple strongly to the third generation and weakly to the first two
generations, and vice versa.

For the charged-current interactions in the quark sector, one observes that
in the case of ignoring the new physics effect, quark mixing is described by
a unitary matrix $V=L_u^{\dagger }L_d$ which is identified as the usual
Cabibbo-Kobayashi-Maskawa (CKM) mixing matrix. With the inclusion of new
physics, the mixing acquires an additional contribution proportional to 
$\sin ^2\phi /x$, with
\begin{equation}
{L_u^{\dagger }}GL_d={L_u^{\dagger }}L_dL_d^{\dagger }GL_d=VL_d^{\dagger
}GL_d={L_u^{\dagger }}GL_uV.
\end{equation}
Therefore, we would expect the extracted values of the CKM matrix elements
to be slightly modified due to the new contributions of the model.

In this model, 
lepton mixing is an exciting possibility. 
Needless to say, there are already significant constraints on lepton
universality and lepton number violation from the low energy data.
As an example is the almost vanishing decay width 
$\Gamma _{\mu^- \rightarrow e^- e^+ e^-} $ which severely 
suppresses any possible mixing between the first
and second lepton generations. 
Similarly, the experimental limit on the decay 
width $\Gamma_{\mu^- \rightarrow e^- \gamma}$ does not favor 
such a mixing. 
Since the other
lepton number violation processes, especially those involving the third
family, are not as well constrained as $\mu \rightarrow eee$ and 
$\mu \rightarrow e \ \gamma $ \cite{data},
it is still interesting to explore such a possibility.
Furthermore, FCNCs can exist in the neutrino sector 
in spite that the neutrinos are massless, 
that may induce an interesting effect to the neutrino 
oscillation phenomena.
As to be shown later, FCNCs are unavoidable in the quark sector
of the model, which can lead to appreciable effects that
can be verified or ruled out by future data on Kaon and $B$ 
physics.

In the following sections, we discuss the effect 
of the new physics predicted by this model to low energy experiments,
and derive the constraints on the parameter space of the model
from the present data. 
Using the latest LEP/SLC data we update 
our previous analysis in Ref.~\cite{ehab5}. For completeness, we also   
study the constraints from current data on a model in which only the  
top and bottom doublet has a different $SU(2)$ gauge interaction, 
which is another possible model of top quark interactions. 
Furthermore, we shall systematically include all the low energy data  
from Tau, Kaon, and $B$ physics, and identify a few  
interesting observables that can be sensitive to this type of  
new physics.  
We have also examined the one-loop contribution to the 
$K^0$-$\overline{K^0}$, $B^0$-$\overline{B^0}$ mixing, and to the  
branching ratio of $b\rightarrow s \gamma$. 

\section{Constraints Imposed by $Z$-Pole Data}
\indent

In the SM, the parameters $\alpha$, $G_F$, and $M_Z$ are determined 
through three experimental measurements, e.g.,
$e$-$p$ scattering, $\mu $ decay, and $Z$ peak at LEP/SLC,
respectively.
In this model, two additional parameters enter through the gauge sector.
These two parameters can be taken as $x$ and $\sin^2\phi $ (or equivalently, 
the heavy
gauge boson mass $M_{Z^\prime }$ and its decay width $\Gamma_{Z^\prime}$). 
Similar to the SM case, it is necessary to fix the input 
parameters $\alpha $, 
$G_F$, $M_Z$, $\sin^2\phi $, and $x$ in this model to make prediction and 
compare with experimental data.
The first three parameters can be fixed in the same way as the SM,
and the last two parameters, $\sin^2\phi $ and $x$, 
will be constrained through available data. 
Because of the symmetry-breaking pattern, the electromagnetic
coupling $\alpha$ coincides with the SM value.
To fix the weak coupling constant, we use the $\mu$-lifetime to define 
$G_F$. We calculate the $\mu $-decay width in this
model by including the $W$ and $W^{\prime }$ contributions. We find that, 
as to be discussed later, 
$G_F=G_F^{{\rm {SM}}}$ (equivalently $v=v^{{\rm {SM}}}$) as long as one
demands no mixing between the first and second lepton families \cite{ehab5}.
Finally, we define $M_Z$ using the $Z$ peak at LEP/SLC, i.e., 
$M_Z=M_Z^{\rm {SM}}$. 

In Ref.~\cite{ehab5} we studied the constraints
imposed by the already existed LEP and SLC data,
we found that the lower
bound on the heavy gauge boson mass was $M_{Z^\prime}\simeq 1.3$ TeV at
the $3\sigma$ level. The lower limit on $M_{Z^\prime}$ was
established for small values of $\sin^2\phi$, for larger values of 
$\sin^2\phi$ the lower bound on $M_{Z^\prime}$ is larger.
Since the $Z$-pole physics program at LEP has completed, it is worthwhile
to update our previous analysis using the most recent data.
Following Ref.~\cite{ehab5}, we calculate the changes in the 
relevant physical 
observables relative to their SM values to leading order in $1/x$, i.e.,
\begin{equation}
O=O^{\rm{SM}}\left(1+\delta O\right ) \, ,
\end{equation}      
where $O^{\rm{SM}}$ is the SM prediction (including the one-loop 
SM correction) for the observable $O$, and $\delta O$ 
represents the new physics effect to leading order in $1/x$.
We list the calculated observables as follows:
\begin{eqnarray*}
\Gamma_Z = \Gamma_Z^{\rm{SM}}\left( 1+\frac{1}{x}\left[-0.896\sin^4\phi 
               + 0.588\sin^2\phi\right]\right)\, ,
\end{eqnarray*}
\begin{eqnarray*}
R_e =R_e^{\rm{SM}}\left(1 + \frac{1}{x}\left[0.0794\sin^4\phi + 
           0.549\sin^2\phi\right]\right)\, ,
\end{eqnarray*}
\begin{eqnarray*}
R_\mu =R_\mu^{\rm{SM}}\left(1 + \frac{1}{x}\left[0.0794\sin^4\phi +
           0.549\sin^2\phi -2.139\sin^2\beta 
\sin^2\phi\right]\right)\, ,
\end{eqnarray*}
\begin{eqnarray*}
R_\tau =R_\tau^{\rm{SM}}\left(1 + \frac{1}{x}\left[0.0794\sin^4\phi +
           0.549\sin^2\phi -2.139\cos^2\beta 
\sin^2\phi\right]\right)\, ,
\end{eqnarray*}
\begin{eqnarray*}
A_{FB}^e={(A^e_{FB}})^{\rm{SM}}\left(1+
\frac{1}{x}\left[10.44\sin^4\phi\right]
          \right)\, ,
\end{eqnarray*}
\begin{eqnarray*}
A_{FB}^\mu={(A^\mu_{FB})}^{\rm{SM}}
\left(1+\frac{1}{x}\left[10.44\sin^4\phi +
               12.14\sin^2\beta\sin^2\phi\right]\right)\, ,
\end{eqnarray*}
\begin{eqnarray*}
A_{FB}^\tau={(A^\tau_{FB})}^{\rm{SM}}\left(1+\frac{1}{x}\left[10.44    
    \sin^4\phi +12.14\cos^2\beta\sin^2\phi\right]\right)\, ,
\end{eqnarray*}
\begin{eqnarray*}
A_e =A_e^{\rm{SM}}\left(1+\frac{1}{x}\left[ 
5.22\sin^4\phi\right]\right)\, ,
\end{eqnarray*}
\begin{eqnarray*}
A_\tau =A_\tau^{\rm{SM}}\left(1+\frac{1}{x}\left[ 5.22\sin^4\phi
                      +12.14\cos^2\beta\sin^2\phi\right]\right)\, ,
\end{eqnarray*}
\begin{eqnarray*}
\sigma_h^0 ={(\sigma_h^0)}^{\rm{SM}} 
\left(1+\frac{1}{x}\left[ -0.01 \sin^4\phi
            -0.628\sin^2\phi\right]\right)\, ,
\end{eqnarray*}
\begin{eqnarray*}
M_W = M_W^{\rm{SM}}\left(1+\frac{1}{x}
\left[1+0.215\sin^4\phi\right]\right)\, ,
\end{eqnarray*}
\begin{eqnarray*}
R_b=R_b^{\rm{SM}}\left(1+\frac{1}{x}\left[ -0.015\sin^4\phi
                          + 1.739\sin^2\phi\right]\right)\, ,
\end{eqnarray*}
\begin{eqnarray*}
R_c=R_c^{\rm{SM}}\left(1+\frac{1}{x}\left[0.038\sin^4\phi-
                           0.549\sin^2\phi\right]\right)\, ,
\end{eqnarray*}
\begin{eqnarray*}
A_b=A_b^{\rm{SM}}\left(1+\frac{1}{x}\left[0.068\sin^4\phi
                          + 0.157\sin^2\phi\right]\right)\, ,
\end{eqnarray*}
\begin{eqnarray*}
A_c=A_c^{\rm{SM}}\left(1+\frac{1}{x}\left[0.514\sin^4\phi\right]\right)\, ,
\end{eqnarray*}
\beq
A_{FB}^c={(A_{FB}^c)}^{\rm{SM}}\left(1+\frac{1}{x}
           \left[5.734\sin^4\phi\right]\right)\, ,
\enq
where $\beta$ is the lepton mixing angle, which will be discussed in the
following sections. 
In this analysis we do not include the measurement of 
$A_{LR}$ at SLC and the
measurement of $A_{FB}^b$ at LEP. 
The quantity $A_{LR}$ in the proposed model is identical to 
$A_e$, therefore, this model cannot explain the discrepancy 
between the SLC measurement 
$A_{LR}=0.1547\pm 0.0032$ and the LEP measurement $A_e=0.1399\pm 0.0073$
\cite{lep}. 
The SM predicts $A_{FB}^b=0.1040$, which is more than 
$2\sigma$ 
above the LEP measurement $A_{FB}^b=0.0984\pm 0.0024$ \cite{lep}. 
The new contribution in this model to $A_{FB}^b$ can be found as
\beq
A_{FB}^b={(A_{FB}^b)}^{\rm{SM}}\left(1+ \frac{1}{x}\left[
           5.287\sin^4\phi + 0.157\sin^2\phi\right]\right)\, .
\enq
which is positive and thus it worsens the discrepancy with LEP data.
Therefore, we cannot accommodate either 
$A_{LR}$ or $A_{FB}^b$ in this model at the $2\sigma$ level.

Following Ref.~\cite{ehab5} and using the most recent LEP and SLC
measurements \cite{lep}, shown in Table 1 (which includes
the total width of the $Z$
boson $\Gamma_Z$, $R_e$, $R_\mu$, $R_\tau$, 
the vector $g_{Ve}$ and the 
axial-vector $g_{Ae}$ couplings of the electron, the ratios 
$g_{V(\mu,\tau)}/g_{Ve}$, $g_{A(\mu,\tau)}/g_{Ae}$, 
$A_{FB}^e$, $A_{FB}^\mu$, $A_{FB}^\tau$, $A_e$, $A_\tau$, $M_W$, 
the hadronic cross section $\sigma_h^0$, $R_b$, $R_c$, $A_b$, $A_c$, and 
$A_{FB}^c$), we update the allowed values of $\sin^2\phi$ and $x$ at 
the $2\sigma$ level. The SM prediction for the observables 
listed in Table 1 is given for $m_t=175$ GeV, $\alpha_s=0.118$, 
$m_H=100$ GeV, $1/{\alpha(M_Z^2)}=128.75$,
$M_Z=91.187$ GeV, and $G_F=1.16637 \times 10^{-5} \, {\rm GeV}^{-2}$
\cite{hagiwara}.

In Figure 1 (solid curve) we show the minimal 
$Z^\prime$ mass as a function of $\sin^2\phi$
at the $2\sigma$ level for the case that there is no mixing in the 
lepton sector. 
We find that $M_{Z^\prime}$ is constrained to be larger than 
about $1.9$ TeV.
(At the $3\sigma$ level, this corresponds to about 1.4 TeV.)
In Figure 2 (solid curve), we show the constraint for the quantity $x$ as a 
function of $\sin^2\phi$. We find that $x$ can be as small as 20 for 
the smallest 
value of $\sin^2\phi$ ($=0.04$), and it increases as $\sin^2\phi$ 
increases. For example,  $x>90$ for $\sin^2\phi>0.2$.
Furthermore, the quantity $\sin^2\phi/x$ is constrained by data to be 
less than about $2\times 10^{-3}$ for a large range of $\sin^2\phi$.

We find 
the most important factor in constraining the free 
parameters of the model is the breakdown of the universality property
of the the gauge boson couplings to leptons. 
The $Z$-pole observable that imposes the most stringent constraint 
on the model is $R_\tau$, which is the ratio of the partial decay widths of 
$Z\rightarrow \tau^+ \tau^-$ and the hadronic modes.
The measurement of $\Gamma_Z$ also plays 
an important role secondary to $R_\tau$, especially for small $\sin^2\phi$,
due to the high
precision of data. It is interesting to note that, as shown in
Figure 1 (dotted curve), without including the leptonic observables 
from the $Z$-pole data, i.e., only including $M_W$, $R_b$, $R_c$, $A_b$, $A_c$,
and $A_{FB}^c$,
 the bounds on $M_{Z^\prime}$ is about 900 GeV at the $2\sigma$ level.
Also, Figure 2 (dotted curve), shows that  
$x>5$ for $\sin^2\phi=0.04$ and $x>24$ for $\sin^2\phi>0.2$. In this case,
the important constraint is coming from the measurement of $R_b$.
The last bound is relevant for models in which only the top and bottom 
doublet has a different $SU(2)$ gauge interaction.
In Table 1, we also show the predictions of this model for the $Z$-pole 
observables with three choices of the parameters $x$, $\sin^2\phi$, and 
$\sin^2\beta$.
          
\begin{figure}[t]
\begin{center}
\leavevmode\psfig{file=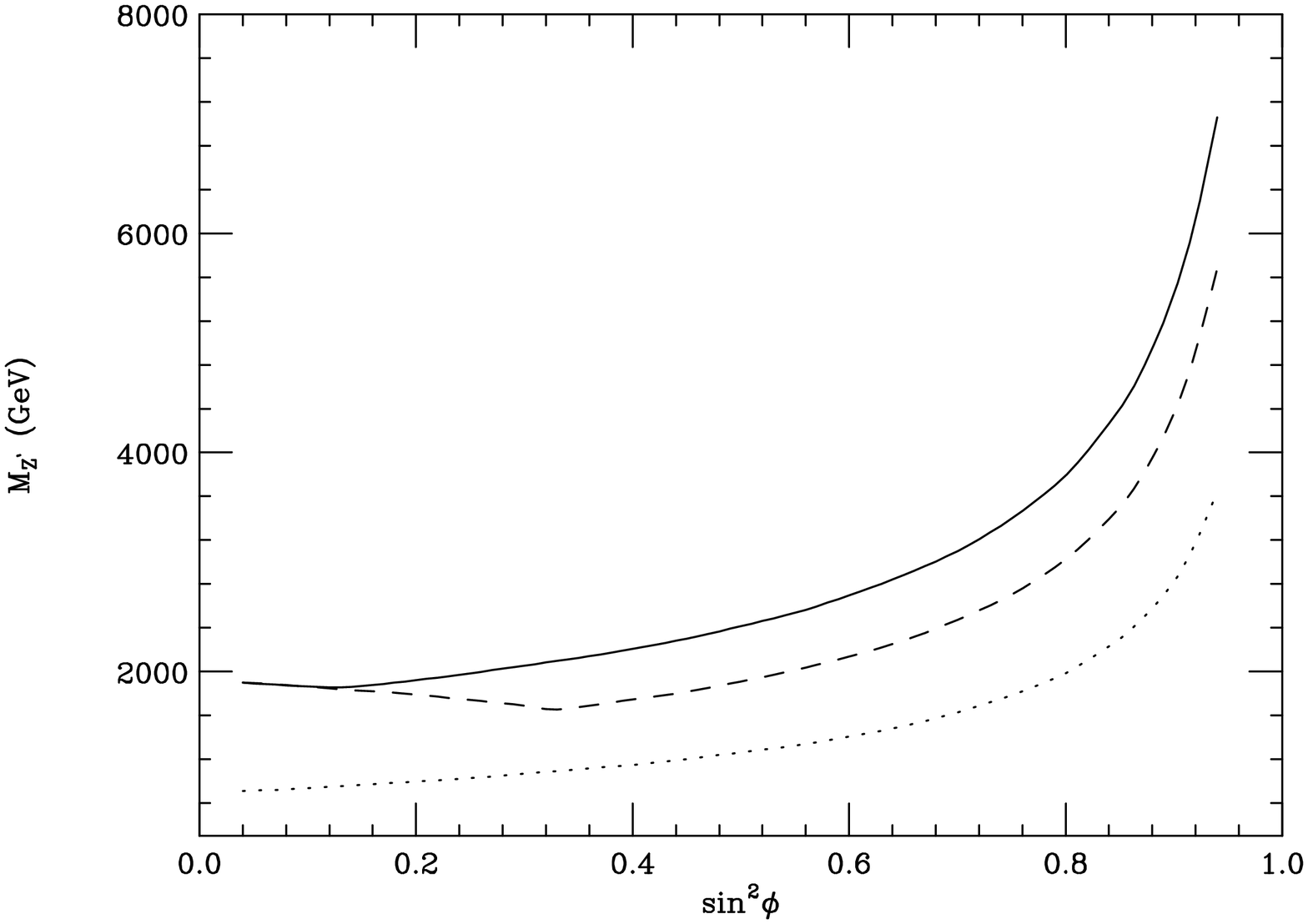,angle=90,height=8.cm}
\end{center}
\caption{The lower bound on the heavy $Z^\prime$ mass as a function of 
 $\sin^2\phi$ at the $2\sigma$ level. 
     Solid curve: including all $Z$-pole data and assuming no lepton mixing.
     Dashed curve: including all $Z$-pole data and assuming maximal lepton 
     mixing ($\sin^2\beta=0.5$).
     Dotted curve: only including the hadronic measurements in the fit and 
     assuming no lepton mixing.}
\label{fig1}
\end{figure}
                                    
\begin{figure}[t]
\begin{center}
\leavevmode\psfig{file=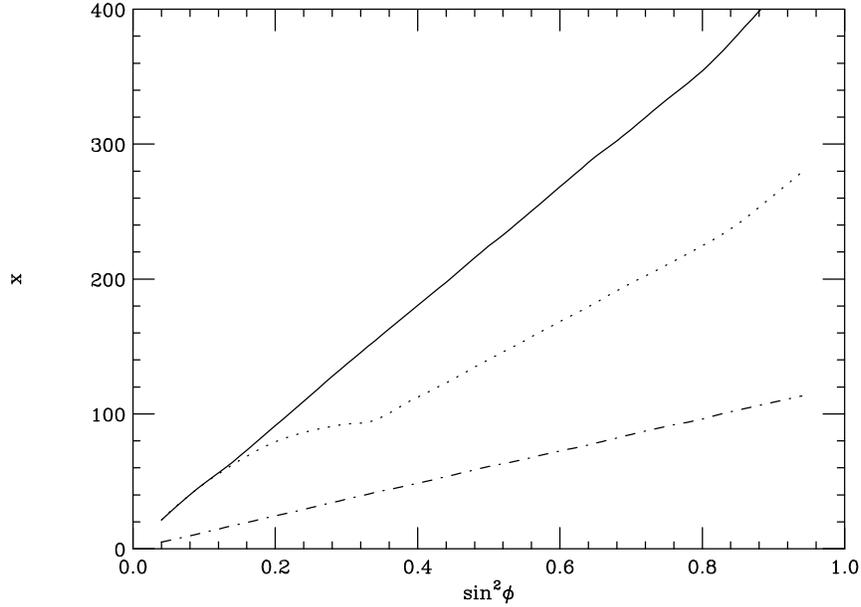,angle=90,height=8.cm}
\end{center}
\caption{The lower bound on the parameter $x$ as a function of 
 $\sin^2\phi$ for the $2\sigma$ level.
     Solid curve: including all $Z$-pole data and assuming no lepton mixing. 
     Dashed curve: including all $Z$-pole data and assuming maximal lepton 
     mixing ($\sin^2\beta=0.5$).
     Dotted curve: only including the hadronic measurements in the fit and 
     assuming no lepton mixing.}
\label{fig2}
\end{figure}
\newpage

\section{Low-Energy Constraints}

\indent

Even though the $Z$-pole data already impose significant constraints, 
this model has a rich structure that can be 
further examined at much lower energy scales. 
In the following sections, 
we would like to examine those constraints obtained from 
 the low-energy hadronic, leptonic, and semi-leptonic data. 
We will concentrate on the very low-energy regime, i.e., physics at
zero-momentum transfer, and examine whether 
the parameters of the model can be better constrained 
than those imposed by LEP and SLC data. 

To study the low-energy region, it is necessary to understand the form of the 
four-fermion current-current
interaction at zero-momentum transfer. The four-fermion
charged-current weak interactions are given by \cite{ehab5,ununi} 
\begin{equation}
\frac 2{v^2}(j_l^{\pm }+j_h^{\pm })^2+\frac 2{u^2}j_h^{+}\,j_h^{-}\,.
\label{cc}
\end{equation}
The first term refers to the SM contribution,
 while the second term expresses
the new contribution to the order $1/u^2$. The charged current
$j_l^{\pm } $ refers to the first two fermion generations, while 
$j_h^\pm $ refers to the third generation. For example, for the lepton sector,
$ j_h^{+}=\overline{\tau _L}\ \gamma _\mu \ \nu _{\tau _L}$. 
We note that in the above formula, the
charged currents $j_h^{\pm }$ are written in terms of the weak eigenstates 
$\tau _L$ and $\nu _{\tau _L}$ and not the mass eigenstates.

Similarly, the neutral-current four-fermion interactions are given by
\cite{ehab5, ununi} 
\begin{equation}
\frac 4{v^2}(j_l^3+j_h^3-\sin ^2\theta \ j_{{\rm {em}}})^2+\frac
4{u^2}(j_h^3-\sin ^2\phi \sin ^2\theta \ j_{{\rm {em}}})^2\,,  \label{nc}
\end{equation}
where, $j_{l,h}^3$ refers to the left-handed $T^3_{l,h}$ currents, while 
$j_{\rm {em}}$ represents the full electromagnetic current of 
the three families.
The first term refers to the SM contribution while the second one represents
the extra contribution. For example, for the
lepton sector,
\begin{equation}
j_h^3=\overline{\tau _L}\ \gamma _\mu \left( \frac{-1}2\right) \tau _L\ 
+ 
\overline{\nu _{\tau L}}\ \gamma _\mu \left( \frac 12\right) \nu _{\tau _L},
\end{equation}
and
\begin{equation}
j_{{\rm em}}=
\overline{e}\ \gamma _\mu (-1) e \ 
+ 
\overline{\mu}\ \gamma _\mu (-1) \mu \ 
+
\overline{\tau}\ \gamma _\mu (-1) \tau \, ,
\end{equation}
in terms of the weak eigenstates.

For clarity, we shall separately discuss below the effects from the 
lepton and quark sectors to the lepton number violation phenomena,
as well as the kaon and bottom physics.

\section{The Lepton Sector}

\indent

As previously mentioned, lepton mixing is an interesting feature of this
model. However, because of the almost null measurement of 
$\mu^{-}\rightarrow e^{-}e^{+}e^{-}$ and $\mu^- \to e^- \gamma$,
we expect the  mixing between the 
first and second lepton families to be highly suppressed. 
Nevertheless, fermion mixing may be large 
between the third family and 
the first or second family. To clarify this point, we write the unitary
matrix $L_e$, which is introduced to diagonalize the mass matrix 
$M_e$, in the general form 
\begin{equation}
L_e={\pmatrix{L_{11} & L_{12} & 
L_{13} \cr L_{21} & L_{22} & L_{23} \cr L_{31} & L_{32} & L_{33} \cr}} \,.
\end{equation}
It is easy to show that 
\begin{equation}
L_e^{\dagger }GL_e=
{\pmatrix{|L_{31}|^2 & L_{31}^\ast L_{32} & L_{31}^\ast L_{33} \cr 
              L_{31}L_{32}^\ast & |L_{32}|^2 & L_{32}^\ast L_{33} \cr 
              L_{31}L_{33}^\ast &  L_{32}L_{33}^\ast & |L_{33}|^2 \cr}}\,.
\label{LG}
\end{equation}
Thus, leptonic FCNC dynamics only depends on the third row of the
mixing matrix $L_e$. In other words, we can only probe the third row of the
unitary matrix $L_e$ through the leptonic FCNC processes.

Using the expression for the four-fermion neutral-current interaction, a direct
calculation of the decay width $\mu \longrightarrow $ $eee$ yields 
\begin{equation}
{\rm{Br}} (\mu \rightarrow eee)=
 \frac{|L_{31}|^2|L_{32}|^2}{4x^2}\left( 
(|L_{31}|^2-2\sin ^2\phi\sin ^2\theta)^2
 +4\sin ^4\phi \sin ^4\theta \right) .
\end{equation}
Notice that the partial decay width of $\mu \to eee$
is already of order $1/x^2$. Therefore, 
to keep the leading contribution of order $1/x^2$, 
we set the total decay width, used in the above equation, to be the SM value.
The above branching ratio has to be compared with the very stringent limit
set by data 
which is less than $10^{-12}$ \cite{data}. Thus, a severe constraint on 
the following combination is established:
\begin{equation}
 \frac{|L_{31}|^2 |L_{32}|^2 \sin^4\phi}{x^2} \lsim 1.6\times 10^{-11}\, .
\end{equation}
As shown in the previous section, the $Z$-pole observables bound the 
quantity of $\sin^2\phi/x$
to be less than 
$2\times 10^{-3}$. 
Therefore, taking $\sin^2\phi/x \sim 2\times 10^{-3}$, 
we get  
\begin{equation}
 {|L_{31}|}^2 {|L_{32}|}^2 < 4\times 10^{-6}\, .
\end{equation}

Another process to consider is $\mu\rightarrow e \ \gamma$, which 
can only occur via loop correction in this model.
The experimental limit on this branching ratio is found to be less than
$4.9\times 10^{-11}$ \cite{data}. 
A one-loop calculation of the branching ratio in the model yields
\begin{equation}
{\rm{Br}}(\mu \rightarrow e \ \gamma )\simeq 8.7\times 10^{-4}
\frac{|L_{31}|^2 |L_{32}|^2}{x^2}(1+1.2\sin ^2\phi +
1.2\sin ^4\phi )\, ,
\end{equation}
which implies
\begin{equation}
 \frac{|L_{31}|^2|L_{32}|^2}{x^2} < 5.6 \times 10^{-8} \, 
\end{equation}
when compared with data.
For the smallest possible value of $x$($\sim 20$) allowed by $Z$-pole 
data, the above constraint yields
\begin{equation}
 |L_{31}|^2|L_{32}|^2 \lsim 2.2 \times 10^{-5}\, ,
\end{equation}
which is weaker (by a factor of 5) than the one imposed by 
the measurement of $\mu\rightarrow eee$.
Other limits on FCNC processes, such as $\tau \rightarrow eee$, 
$\tau \rightarrow \mu\mu\mu$, $\tau \rightarrow ee\mu$, are not as severe 
as the ones mentioned above. (Their branching ratios are typically bounded
from above at the order of $10^{-6}$ \cite{data}.)

The above constraints on the elements of the lepton mixing matrix
$L_e$ can be automatically satisfied if $L_{31}=0$ and/or $L_{32}=0$,
which means there is no mixing between the third family and 
the first and/or the second family leptons. Consequently, with this choice, 
this model predicts no transition between $\mu$ and $e$ leptons.
Although both cases of lepton mixing are allowed, it is
more natural to assume the mixing strength between leptons to be directly
related to their masses. If so, one would expect the mixing between the
second and third families to be more significant than the first and third
families. 
Hence, in the following discussion, we will assume that leptonic mixing
is only allowed between the second and third families
 (i.e., we set $L_{31}=0$). 

The lepton-mixing
matrix has the form $L_e^{\dagger }GL_e$, given in Eq.~(\ref{LG}), 
where the
matrix $G$ is defined in Eq.~(\ref{G}). Using unitarity of $L_e$ and 
taking $L_{31}=0$,
we have $|L_{32}|^2+$ $|L_{33}|^2=1$. Therefore, the mixing
matrix between the second and third lepton families can be simply expressed
in terms of a one free real parameter, and 
the $2\times 2$ mixing matrix can be
written as 
\begin{equation}
\pmatrix{\sin^2\beta & \cos\beta\sin\beta \cr \cos\beta\sin\beta & \cos^2
\beta \cr}\,,
\end{equation}
where $\sin \beta $ is a free parameter of the model for describing the
mixing between the second and third lepton families. The phases in the
matrix $L_e^{\dagger }GL_e$ can be simply absorbed in the definitions of the
lepton fields.
It is easy to see that if
there is no mixing among leptons, then all the leptonic decay rates
are identical to the SM, and $\tau $ lifetime is not modified. 
(This also explains why $G_F=G_F^{\rm {SM}}$
from the $\mu $-decay if there is no mixing between the first and second
lepton families.) If the lepton mixing involves the third family,
then the lifetime of the $\tau $ lepton will be modified.

At this stage it is relevant to return back to the LEP and SLC 
data and study the
new constraints on the model if a mixing between $\mu$ and $\tau$ is allowed. 
In this case we also need to include the limit on the branching ratio of
$Z\rightarrow \mu \tau$, which is found to 
be less than $1.7\times 10^{-5}$ at the $2\sigma$ 
level \cite{data}. 
In Figure 1 (dashed curve), we depict the new constraints on 
$\sin^2\phi$ and $M_{Z^\prime}$ 
for the case of a maximal possible mixing, i.e., 
$\sin^2\beta=0.5$. We find that the lower limit on the heavy mass is
 reduced to $M_{Z^\prime}\approx 1.7$ TeV, which is slightly lower than 
that for the case of no mixing ($\approx 1.9$ TeV).
The reason for this lower bound is due to the reduced non-universal 
effect in $R_\tau$. 
In Figure 2 (dashed curve) we show the new constraint on $x$, assuming 
the maximal lepton mixing.
We find that for the smallest value of 
$\sin^2\phi=0.04$, the value of $x$ can be as low as 20.
For $\sin^2\phi=0.2$, $x > 80$. 
The quantity $\sin^2\phi/x$ is found to be less than about $0.3\%$.
It is interesting to notice that the lower bound $x=20$ is the same 
for both cases of $\sin^2\beta=0$ and $\sin^2\beta=0.5$. The reason is that 
for small values of $\sin^2\phi<0.2$, the measurement $\Gamma_Z$, which is 
independent of the mixing angle $\sin^2\beta$, plays 
the important rule in constraining the parameter $x$. 
In Table 1, we give a few predictions of this model with various
$\sin^2\beta$ for the $Z$-pole observables.

Next, we examine the other interesting low-energy leptonic 
processes and ask 
whether we can learn more about the proposed model. 
We start by examining the decay process 
$\tau \rightarrow \mu \overline{\nu _\mu }\nu_\tau$.
In this model, both the charged and neutral currents contribute 
to the decay width
$\Gamma\left( \tau ^{-}\rightarrow \mu ^{-}\overline{\nu _\mu }\nu _\tau
\right)$. Adding both contributions, we find 
\begin{equation}
\Gamma\left( \tau ^{-}\rightarrow \mu ^{-}\overline{\nu _\mu }\nu _\tau
\right) =\Gamma^{\rm{SM}}\left( \tau ^{-}\rightarrow \mu ^{-}\overline
{\nu _\mu }\nu _\tau \right) \left( 1+\frac{3\cos ^2\beta \sin ^2\beta }
x\right) .
\end{equation}
The only modification to the total decay width, at the order of $1/x$, 
is coming from the partial 
decay width 
$\Gamma\left( \tau ^{-}\rightarrow \mu ^{-}
\overline{\nu _\mu }\nu _\tau\right)$.
The partial decay width 
$\Gamma\left( \tau ^{-}\rightarrow e ^{-}\overline{\nu _e }\nu _\tau\right)$ 
is not modified because of the assumption of no $\tau \leftrightarrow e$ 
mixing. 
The ratio $\Gamma\left(\tau^{-}\rightarrow 
\mu^{-}\overline{\nu _\mu }\nu _\tau\right)/\Gamma\left(\tau^{-}\rightarrow 
e^{-}\overline{\nu _e }\nu _\tau\right)$, 
can be written as
\begin{equation}
\frac{\Gamma\left(\tau^{-}\rightarrow 
\mu^{-}\overline{\nu _\mu }\nu _\tau\right)}{\Gamma\left(\tau^{-}\rightarrow 
e^{-}\overline{\nu _e }\nu _\tau\right)}= 
\frac{{\rm{Br}} \left( \tau ^{-}\rightarrow 
\mu ^{-}\overline{\nu _\mu }\nu _\tau
\right)}
{{\rm{Br}}\left( \tau ^{-}\rightarrow e ^{-}\overline
{\nu _e }\nu _\tau \right)}=f(m_\mu/m_\tau)
\left( 1+\frac{3\cos ^2\beta \sin ^2\beta}{x}\right)\, , 
\end{equation}
where $f(m_\mu/m_\tau)$ is a phase factor given by \cite{park}
\begin{equation}
f(y)=1-8y^2+8y^6-y^8-24y^4\ln(y) \,.
\label{phase}
\end{equation}
Hence, an increase by a 
factor of ${3\cos ^2\beta\sin ^2\beta} /x$
is expected in the above ratio. 

The experimental measurement of ${\Gamma\left(\tau^{-}\rightarrow 
\mu^{-}\overline{\nu _\mu }\nu _\tau\right)}/{\Gamma\left(\tau^{-}\rightarrow 
e^{-}\overline{\nu _e }\nu _\tau\right)}$
 can directly constrain the quantity
$\cos ^2\beta \sin ^2\beta /x$. As shown by the Particle Data Group 
(PDG) \cite{data},
for the ratio $\Gamma\left(\tau^{-}\rightarrow 
\mu^{-}\overline{\nu _\mu }\nu _\tau\right)/\Gamma\left(\tau^{-}\rightarrow 
e^{-}\overline{\nu _e }\nu _\tau\right)$,
the average of the available experimental data 
yields $0.978\pm 0.011$ while the result of a global fit gives
$0.976 \pm 0.006$.     
In this model, new decay channels for the $\tau$ lepton can occur, e.g., 
$\tau\rightarrow \mu \mu \mu$ and $\tau \rightarrow \mu \gamma$. 
However, as to be discussed below,  
their decay widths can only be modified at the order $1/x^2$.
Thus, to the order $1/x$, we can use both data, the average and 
the fit results, to constrain the parameter $x$.
Using the PDG data average and assuming a maximal lepton mixing 
$\sin^2\beta=0.5$, we find $x> 27$ at 
the $2\sigma$ confidence level. On the other hand,
using the PDG fit result we find $x>48$. The difference in the $x$ range, 
27 to 48 can then be interpreted as the theoretical error in our model. 

For a lepton mixing angle $\sin^2\beta$ smaller than 0.5,
the constraint on the parameter $x$ is more relaxed. 
If there is no lepton mixing at all, then 
the decay width of 
$\tau ^{-}\rightarrow \mu ^{-}\overline{\nu _\mu }\nu _\tau $
is not modified as compared with the SM prediction.
Since this decay width is independent of the
parameter $\sin^2\phi $ (gauge coupling) and the only dependence besides
lepton mixing is the parameter $x$ (the ratio of the two
symmetry-breaking scales of the gauge group), 
this measurement imposes a direct and significant constraint on $x$
for a non-vanishing $\sin^2\beta$.

Another interesting process for testing this model is to detect
$\tau \rightarrow \mu \mu \mu $. One can show
that\footnote{
Our prediction for ${\rm{Br}}(\tau \rightarrow \mu \mu \mu )$ is slightly  
different from that in Ref.~\cite{lee}.},
keeping the leading contribution in $1/x$,
\begin{equation}
\frac{{\rm{Br}}(\tau^- \rightarrow \mu^- \mu^- \mu^+)}
{{\rm{Br}} (\tau^- \rightarrow \mu^- \nu_\tau {\overline{\nu }}_\mu )}=
\frac{\sin ^2\beta \cos ^2\beta }{4x^2}\left( \sin ^4\beta -4\sin ^2\beta \sin
^2\theta \sin ^2\phi +8\sin ^4\phi \sin ^4\theta \right) \,\label{muuu}\,.
\end{equation}
This decay width will also impose a direct constraint on the parameters of
the model. For $\sin^2\beta=0.5$ and 
$\sin^2 \phi=0.04$, the predicted branching ratio is 
\begin{equation}
{\rm{Br}}(\tau ^{-}\rightarrow \mu ^{-}\mu ^{-}\mu ^{+})\simeq
\frac{0.0025}{x^2}\, . 
\end{equation}
If we compare this effect with data which is found to be less than 
$1.9\times 10^{-6}$ \cite{data}, the parameter 
$x$ is constrained to be above 37,
which is consistent with the constraint ($x>27\sim 48$) derived from the 
measurement of 
${\rm{Br}}(\tau\rightarrow \mu \overline{\nu _\mu }\nu _\tau)/
{\rm{Br}}(\tau\rightarrow e \overline{\nu _e }\nu _\tau)$. 

Other processes to consider is the lepton number violation
process $\tau \rightarrow \mu \, \gamma $, which can only occur
in this model at the loop level.
For this process, up to the order $1/x$, there are
four diagrams which contribute to the one-loop amplitude, Two
of those diagrams involve either two $W$ or two $W^\prime$ exchange. The other
two diagrams involve $Z$ or $Z^\prime$ exchange (due to FCNC). A
detailed calculation of the branching ratio yields
\begin{equation}
{\rm{Br}}(\tau \rightarrow \mu \gamma )\simeq 1.5\times 10^{-4}
\frac{\sin^2\beta \cos ^2\beta }{x^2}(1+1.2\sin ^2\phi +1.2\sin ^4\phi )\, .
\end{equation}
This  result has to be compared with the limit imposed by 
data (less than $4.2\times 10^{-6}$ \cite{data}). 
For a maximal possible mixing effect, 
the present limit on the above branching ratio is not of any 
significance in constraining the values of $x$ ($>3$).

The final leptonic observable we consider is the anomalous magnetic dipole 
moment of the muon, $a_\mu=\frac{1}{2}{(g-2)}_\mu$. A precise measurement 
of $a_\mu$ is underway at Brookhaven National Laboratory (BNL) with a 
perspective goal \cite{bnl_amu} of 
\begin{equation}
\Delta a_\mu^{\rm{exp.}} = 4.0\times 10^{-10}\,\,.
\end{equation}
At this level of accuracy, higher order electroweak corrections become important 
and new physics at higher energy scales can be probed.

The one-loop electroweak contribution to $a_\mu$ as predicted by the SM is  \cite{amu_sm}
\begin{equation}
a_\mu^{\rm{weak}} \approx 19.5\times 10^{-10}\, \, .
\end{equation} 
In our proposed model the one-loop electroweak contribution is modified due to 
the modified couplings and 
the new heavy gauge bosons. We calculate the new contribution to $a_\mu$ at 
the one-loop level. 
We find the new contribution to the anomalous magnetic dipole moment to be
\begin{equation}
a_\mu^{\rm{new}}\approx a_\mu^{\rm{weak}} \frac{\sin^2\beta}{x}\, \, . 
\end{equation}
Using the $Z$-pole constraints, for a maximal mixing, $\sin^2\beta=0.5$ 
and $x\geq 20$ we conclude that the new contribution does not exceed 
the level of $0.5 \times 10^{-10}$. 
Therefore, the predicted new effect to $a_\mu$ is too small to be detected even 
at the perspective precision at BNL.

In conclusion, assuming the third family lepton does not mix with the 
first family lepton, then the partial decay widths of
$\mu \rightarrow eee$, $\mu \rightarrow e \gamma$, and
$\tau \rightarrow e \overline{\nu _e }\nu _\tau $  
are not modified.  
However, for the maximal mixing case, 
the measurement of the ratio
$\Gamma(\tau \rightarrow \mu \overline{\nu _\mu }\nu _\tau)/
 \Gamma(\tau \rightarrow e \overline{\nu _e }\nu _\tau)$ constrain 
the parameter $x>27\sim 48$.  
Also the lepton number violation
process $\tau \rightarrow \mu \mu \mu $  
provides the constraint $x>37$ consistent with the above measurement. 
Therefore, the above two measurements give a stronger constraint than the  
$Z$-pole data for $\sin^2\phi <0.1$ (cf. Figure 2).
On the other hand, given the current experimental data,
the decay process $\tau\rightarrow \mu \gamma$ and the anomalous magnetic dipole
moment of the muon have not yet played a 
significant role in constraining the model. 
Nevertheless, if the above discussed processes can 
be measured to a better accuracy 
in future experiments, they can further test the proposed model.
In discussing the predictions of our model to other 
low-energy processes 
we will use the range $x\geq 20$ for $\sin^2\beta=0.0$, and
$x\geq 48$ for $\sin^2\beta=0.5$.  

\section{The Quark Sector}

\indent

The quark sector has a far more rich structure than the lepton sector
in this model. To completely describe the interactions of 
gauge bosons and quarks, it requires two 
mixing matrices $L_u$ and $L_d$ because both the up- and down-type 
quarks are massive. 
As noted in Eq.~(\ref{qu2}) 
the neutral-current mixing matrices 
($L_u^{\dagger }GL_u$ and $L_d^{\dagger }GL_d$)
are related to the charged-current mixing matrix ($L_u^{\dagger }L_d$).
Because of the experimental evidence of the CKM matrix in charged 
currents, FCNCs must occur in 
the interaction of quarks to gauge bosons.

First we make the following observation. 
Assume neither up- or down-type quark sectors has FCNC, 
i.e., assume $L_u$ (and $L_d$) has the general form
\begin{equation}
L_u=\left( 
\begin{array}{ccc}
u_{11} & u_{12} & 0 \\ 
u_{21} & u_{22} & 0 \\ 
0 & 0 & u_{33}
\end{array}
\right) .
\end{equation}
It is straight forward to show that 
$L_u^{\dagger }GL_u=L_d^{\dagger }GL_d=G$ and that the charged-current mixing
matrix $V=L_u^{\dagger }L_d$ has the same general form as $L_u$ and $L_d$.
This means that $V$ will only mix the first and second generation, i.e., the
CKM matrix is a $2\times 2$ matrix. 
Therefore, unless we assume the existence of FCNCs in the quark
sector, this model cannot explain some observed decay processes,
such as 
 $B_d^0\longrightarrow J/\psi (1S)\, K^0$, in which 
$b\longrightarrow cW^*\longrightarrow c\overline{c}s$ whose
branching ratio was measured to be 
$(7.5\pm 2.1)\times 10^{-4}$ \cite{data}. 
Hence, FCNCs must exist in the quark sector.

Based on the above observation, FCNC data in the quark sector can
be used to further test this model.
FCNCs in the quark sector can be realized 
in three possible ways: (i) in the down-quark sector only, (ii) 
in the up-quark sector only, and (iii) in both sectors. All the
three possibilities have to confront the large body of existing 
low-energy data.
In the following, we investigate these three possibilities, separately.

\subsection{Mixing in the Down-Quark Sector}

\indent

Here, we consider the case that only down-type quarks can mix, 
so that $L_u^{\dagger }GL_u=G$, and $L_d^{\dagger }GL_d=V^{\dagger}GV$.
In this case, the quark interactions to the gauge bosons are given 
in Eq.~(\ref{qu2}) with the above substitutions. 
 Similar to the SM case, the mixing matrix $V$
contains the same number of independent parameters, namely, three real
parameters and one phase. Therefore, there is no extra parameter
in the quark sector in spite of the new features of the model. This implies
that FCNC processes are completely determined by the matrix $V$ in addition
to the other two parameters $\sin^2\phi $ and $x$. The matrix $V^{\dagger }GV$
can be explicitly written as
\begin{equation}
V^{\dagger }GV=\left( 
\begin{array}{ccc}
|V_{td}|^2 & V_{ts}V_{td}^{*} & V_{tb}V_{td}^{*} \\ 
V_{td}V_{ts}^{*} & |V_{ts}|^2 & V_{tb}V_{ts}^{*} \\ 
V_{td}V_{tb}^{*} & V_{ts}V_{tb}^{*} & |V_{tb}|^2
\end{array}
\right) .
\end{equation}
It is interesting to notice that the matrix elements of 
$V^{\dagger }GV$ are naturally small, so that we generally do not expect 
large effects in FCNC processes.

\subsubsection{Charged-Current Phenomenology}

\indent

In general, this model predicts new contributions to 
charged-current processes as well.
Under the down-quark mixing scenario,
the non-standard contribution to charged-current processes can be 
written as
\begin{equation}
\frac{2\sqrt{2}G_F}xj_h^{+}\ j_h^{-} \, ,
\label{eq:cc}
\end{equation}
where \/ $j_h^{-}=\overline{t_L}\gamma _\mu  b_L$, written in terms of the
weak eigenstates $t_L$ and $b_L$. Since we assume no FCNCs in the
up-type quark sector, the top quark does not
mix with the other up-type quarks at tree level.
Furthermore, for the low-energy 
charged-current observables (with momentum transfer 
$q^2$ much less than $M_Z^2$), top quark does not contribute at tree
level. Hence, we conclude that under this scenario, 
no new physics effect to the low-energy charged-current interaction
is expected at tree level.
Therefore, the values of the CKM 
matrix elements extracted from low-energy 
charged-current data coincide with those in the SM.

\subsubsection{Neutral-Current Phenomenology}

\indent

On the contrary,
the neutral-current hadronic and semi-leptonic interactions 
can be modified at tree level for the case of 
down-type quark mixing.
The non-standard contribution to neutral-current processes can be 
written as
\begin{equation}
\frac{4\sqrt{2}G_F}{x} (j_h^3-\sin ^2\phi \sin ^2\theta \ j_{{\rm {em}}})^2\, ,
\end{equation}
where \/ 
$j_h^3$ contains 
$ \overline{b_L}\gamma_\mu (-1/2) b_L$, written in terms of
the weak eigenstate. 
In terms of the mass eigenstates, the following currents are generated: 
$|V_{td}|^2\, \overline{d_L}\gamma_{\mu} d_L$, 
$|V_{ts}|^2\, \overline{s_L}\gamma_{\mu} s_L$, 
$|V_{tb}|^2\, \overline{b_L}\gamma_{\mu} b_L$, 
$V_{td}^* V_{ts}\, \overline{d_L}\gamma_{\mu} s_L$, 
$V_{td}^* V_{tb}\, \overline{d_L}\gamma_{\mu} b_L$, and 
$V^*_{ts}V_{tb}\, \overline{s_L} \gamma_{\mu}b_L$,
whose effects to low-energy FCNC data are discussed as follows.

The first interesting process to investigate is the $K^0$-$\overline{K^0}$ 
mixing, whose transition amplitude 
receives in this model a new contribution at 
the tree level. Up to the order of $1/x$, it is 
\begin{equation}
T=\frac{\sqrt{2}G_F}{x} {(V_{td} V_{ts}^*)}^2 [\overline{s_L}\gamma_\mu d_L]
    [\overline{s_L}\gamma_\mu d_L]\, .
\end{equation} 
In the SM, ignoring the QCD corrections, the short distance 
transition amplitude induced from box diagrams for the 
$K^0$-$\overline{K^0}$ mixing is given by \cite{lashin, inami}
\begin{equation}
T^{\rm{SM}}=\frac{G_F^2 M_W^2}{\pi ^2}
\left[ \lambda_c^2 S(y_c) + \lambda_t^2 S(y_t)+2\lambda_c \lambda_t S(y_c,y_t)
\right] \left[ \overline{s_L}\gamma _\mu \ d_L\right] 
\left[ \overline{s_L}\gamma _\mu \ d_L\right],
\end{equation}
where, $y_c=m_c^2/M_W^2$, $y_t=m_t^2/M_W^2$,  
$\lambda_c= V_{cs}^* V_{cd}$, $\lambda_t= V_{ts}^* V_{td}$, and 
the functions $S(y)$ and $S(y_c,y_t)$ are the Inami-Lim 
functions \cite{inami}:
\begin{eqnarray*}
S(y)=y\left[\frac{1}{4}+\frac{9}{4}\frac{1}{1-y}-\frac{3}{2}
\frac{1}{{(1-y)}^2}\right] -\frac{3}{2} {\left[\frac{y}{1-y}\right]}^3 \ln y
\, \label{s_y} , 
\end{eqnarray*}
\begin{equation}
S(y_c,y_t)=-y_c\ln y_c +y_c\left[\frac{y_t^2-8y_t+4}{4{(1-y_t)}^2}\ln y_t 
   + \frac{3}{4}\frac{y_t}{y_t-1} \right]\, .
\end{equation}
When comparing the non-standard and the SM amplitude, which is 
proportional to the
$\Delta M$ ratio,
we find approximately 
\begin{equation}
\frac{\Delta M}{{\Delta M}^{\rm{SM}}}=
\frac{T}{T^{\rm{SM}}}\lsim \frac{4}{x}\, ,
\end{equation}
in which we have used $m_t=175$ GeV, $m_c=1.5$ GeV, $M_W=80.4$ GeV and
all the CKM elements are taken from Refs.~\cite{data, ali1}.
For $x>20$ (implied by $Z$-pole data), 
it would correspond to a change in the transition amplitude
by less than about 20\%.
Although the mass difference between $K_L$ and $K_S$ states has been 
measured experimentally with a great accuracy (about 0.4\%), 
the theoretical uncertainty in the long distance part of contribution
remains to be improved. 
(Currently, its uncertainty is about $40\%$ to $60\%$ \cite{lashin}.)
To use the $K^0$-$\overline{K^0}$ mixing data to further test this model 
requires a better understanding of the long-distance contribution.

It is well known that the rare decay process
$K^+ \rightarrow \pi^+ \nu\overline{\nu}$ is 
one of the best places to search for new physics. This is because
its decay rate has a small theoretical uncertainty, and 
the long-distance contribution has been estimated to be less than $10^{-3}$
of the short-distance contribution \cite{ellis}.
Recently, E787 collaboration reported the first observation consistent with 
this decay rate and obtained 
${\rm{Br}}(K^+ \rightarrow \pi^+ \nu\overline{\nu})= 4.2^{+9.7}_{-3.5}
\times 10^{-10}$ \cite{e787}. 
The branching ratio predicted by the SM is
${\rm{Br}}^{\rm{SM}}(K^+ \rightarrow \pi^+ \nu\overline{\nu})= 
(9.1\pm 3.8) \times 10^{-11}\,$,
where the error is dominated by the uncertainties of 
the CKM matrix elements \cite{buras}.

Since under the scenario considered, 
this process can occur at the tree level through the 
flavor-changing neutral current 
$s\rightarrow d\, Z \rightarrow d \nu \overline{\nu}$, 
it can be used to test the model.
The expected branching ratio, normalized to 
the predicted branching ratio for $K^+ \rightarrow \pi^0 e^+ \nu_e $, 
can be written as  
\begin{equation}
\frac{{\rm{Br}}(K^+ \rightarrow \pi^+ \nu \overline{\nu})}
{{\rm{Br}}(K^+ \rightarrow \pi^0 e^+ \nu_e)}=
\frac{1}{4x^2} \left( \frac{|V_{td}|^2 |V_{ts}|^2}{|V_{us}|^2}\right)\, .
\end{equation}
It is obvious that the partial decay width of
$K^+ \rightarrow \pi^0 e^+ \nu_e$ predicted by this model coincides with
the SM prediction at tree level for the 
undertaken scenario that the third
family lepton does not mix with the first family lepton.
Therefore, assuming the experimental data  
${\rm{Br}}(K^+ \rightarrow \pi^0  e^+ \nu_e)=
(4.82\pm 0.06)\times 10^{-2}$ \cite{data} to be consistent with the model,
we can compare the predicted 
${\rm{Br}}(K^+\rightarrow \pi^+ \nu\overline{\nu})$ 
with the E787 result.   
After spanning all the allowed  values of the CKM elements, 
 we find that
\begin{equation}
{\rm{Br}}(K^+ \rightarrow \pi^+ \nu\overline{\nu})
\lsim\frac{1.1\times 10^{-7}}{x^2}\, ,
\end{equation}
in which we have included all three neutrino species, i.e.,
$\nu_\mu \overline{\nu_\mu}$, $\nu_\tau \overline{\nu_\tau}$,
$\nu_\mu \overline{\nu_\tau}$, and $\nu_\tau \overline{\nu_\mu}$, so that
the lepton mixing angle dependence cancel. 
Comparing this branching ratio with the E787 result,
we can set a lower bound $ x > 7$  
at the $2\sigma$ level based on one observed event.
For $x>20$ (as implied by $Z$-pole data), this branching ratio
is smaller than about $3\times 10^{-10}$, which is however larger than 
the SM prediction by almost an order of magnitude.

The measurement of ${\rm{Br}}(K^+ \rightarrow \pi^+  \nu \overline{\nu})$
is highly valuable in our analysis because it is independent of the 
parameters
$\sin^2\beta$ and $\sin^2\phi$. It directly constrains the parameter
 $x$ independently of the other parameters.
Hence, an improvement on the measurement of this branching ratio is
very important to test this model.

Similarly, this model predicts non-standard effects for bottom quark 
physics.
The important process to consider is 
$B^0_q$-$\overline{B^0_q}$ mixing where new effect is expected 
to occur at tree level. 
The tree-level transition amplitude is found to be 
\begin{equation}
T=\frac{\sqrt{2}G_F}{x} {({V_{tq}^*} V_{tb})}^2 [\overline{q_L}\gamma_\mu b_L]
    [\overline{q_L}\gamma_\mu b_L]\, .
\end{equation} 
The new contribution can be compared with the SM prediction which is 
given by 
\begin{equation}
T^{\rm{SM}}=\frac{G_F^2 M_W^2}{\pi^2}\left( V_{tq}^{*}V_{tb}\right) ^2
\left[ \overline{q_L}\gamma _\mu \ b_L\right] 
\left[ \overline{q_L}\gamma _\mu \ b_L\right] S(y_t) \label{B_mix},
\end{equation}
where $S(y)$ is given in Eq.~(\ref{s_y}) with $y_t=m^2_t/M^2_W$.
After substituting all the relevant variables by their numerical values,
we find
\begin{equation}
\frac{\Delta M_{B_q}}{(\Delta M_{B_q})^{\rm{SM}}}=
\frac{T}{T^{\rm{SM}}}=\frac{72}{x}\, .
\end{equation}
With the limit on $x$ ($>20$) imposed by the $Z$-pole data alone, 
we expect the new 
contribution to reach the level of $360\%$ for the small possible
values of $x$.
In the case that the third and the second generation fermions 
mix with the maximal strength, the 
${\rm{Br}}(\tau\rightarrow \mu\overline{\nu_\mu} \nu_\tau)$ 
data requires $x>48$, so that  
the new contribution to 
the $B^0_q$-$\overline{B^0_q}$ mixing 
is expected to be less than $150\%$.

The measured value of $\Delta M_{B_d}=0.470\pm 0.019 \, ps^{-1}$ \cite{data} 
can be turned into an information on the CKM elements product 
$|V_{td}V^*_{tb}|$ 
which yields $|V_{td}V^*_{tb}|=0.0084\pm 0.0018$ for the SM \cite{data}. 
In the proposed model, the prediction for $\Delta M_{B_d}$ is larger than 
the SM value by a factor $1+ 72/x$ (adding both SM and the new effect).
Therefore, the extracted $|V_{td} V^*_{tb}|$ will 
be modified accordingly. For example, for $x=20$, we find  
$0.0022< |V_{td}V^*_{tb}| <0.0056$ at the $2\sigma$ level.
This shift is not expected to appreciably affect the unitarity condition 
\cite{ali1}
\begin{equation}
|V_{td}|^2+|V_{ts}|^2+|V_{tb}|^2=0.98\pm 0.30 \, .
\end{equation}
For example, for $x\geq 20$ the deviation from unity will be of the order 
$\sim \frac{72}{x} |V_{td}|^2 \lsim  7\times 10^{-4}\, ,
$
which is much smaller than the present errors.
Also, it is clear that the predicted ratio 
$\Delta M_{B_d}/\Delta M_{B_s}$ in our model is the same as the SM prediction.
Therefore, the extracted ratio $|V_{td}/V_{ts}|$ yields the same SM result.

Next, we consider the CLEO limit on 
${\rm Br}(b\rightarrow s \ell^+\ell^-)$ and study 
its impact on the model. 
At tree level, 
the expected branching ratio is given by
\begin{equation}
\frac{{\rm{Br}}(b\rightarrow s \mu^-\mu^+)}
     {{\rm{Br}}(b \rightarrow c \mu^- \overline{\nu_\mu})}
=\frac{1}{4x^2} 
\frac{|V_{ts}V_{tb}|^2}{|V_{cb}|^2 f(z)}
\left( \sin^4\beta-4\sin^2\beta\sin^2\phi\sin^2\theta +
        8\sin^4\theta\sin^4\phi \right) \, ,
\label{eq:bsuu}
\end{equation}
where $f(z)$ is given in Eq.~(\ref{phase}) and where 
$z=m_c/m_b$ \cite{buras, ahmad}.
Using the experimental data
${\rm{Br}}(b \rightarrow c \mu^- {\overline \nu_\mu})=(10.5 \pm 0.5)\%$ 
\cite{cleo1},
we get
\begin{equation}
{\rm{Br}}(b\rightarrow s \mu^-\mu^+)\lsim \frac{2.1\times 10^{-2}}{x^2}\, 
\end{equation}
after spanning the allowed values of the CKM matrix elements
with $\sin^2\beta=0.5$ and $\sin^2\phi=0.04$.
To agree with the CLEO upper limit, 
which is $5.8\times 10^{-5}$ \cite{cleo2}, $x$ is found to be
larger than 19, which should be compared with the bound ($x>48$)
obtained from the ${\rm{Br}}(\tau\rightarrow \mu \overline{\nu_\mu} \nu_\tau)/
{\rm{Br}}(\tau\rightarrow e \overline{\nu_e} \nu_\tau)$ 
data.
To reach the same sensitivity as $\tau$ decay for $\sin^2\beta=0.5$, 
the measurement of ${\rm{Br}}(b\rightarrow s \mu^-\mu^+)$
has to be improved by a factor of 10, although
in general, they have different dependence on $\sin^2\beta$

With the
assumption that lepton mixing is only present between the third
and the second generation, there is no new contribution to 
the decay rate of $b\rightarrow s e^+ e^-$. 
If we assume the opposite, namely that mixing is
significant between the first and third generation, then we expect the 
decay rate of $b\rightarrow s e^+ e^-$ to dominate the
decay rate of $b\rightarrow s \mu^+ \mu^-$. Since the CLEO bound on 
${\rm{Br}}(b\rightarrow s e^+ e^-)$, 
less than $5.7\times 10^{-5}$ \cite{cleo2},
is similar to the bound
on the $\mu \mu $ channel, 
we expect a similar conclusion on 
constraining the parameter $x$. Furthermore, the decay rate of
$b\rightarrow s e^\pm \mu^\mp$
is highly suppressed because of the sever constraint on
the $e$-$\mu$ mixing 
established from the decay of $\mu \rightarrow eee$ 
and $\mu \rightarrow e \gamma$, as discussed in section 5.  
On the other hand, the branching ratio
${\rm{Br}}(b\rightarrow s \mu^\pm \tau^\mp)$
predicted in this model is of the same order as
${\rm{Br}}(b\rightarrow s \mu^- \mu^+)$.
In the limit of ignoring the mass difference between
$\tau$ and $\mu$, it can be obtained from 
Eq.~(\ref{eq:bsuu}) by multiplying a factor of $2 \cot^2 \beta$ and setting
$\sin^2\phi=0$.
Since this decay mode is absent in the SM, it can be very useful
to further test the model.

Similarly, our model predicts a tree-level contribution to the process 
$b\rightarrow s \nu \overline{\nu}$, whose
branching ratio, when normalized by
${{\rm{Br}}(b \rightarrow c \mu^- \overline{\nu_\mu})}$,
is given as
\begin{equation}
\frac{{\rm{Br}}(b\rightarrow s \nu \overline{\nu})}
     {{\rm{Br}}(b \rightarrow c \mu^- \overline{\nu_\mu})}
=\frac{1}{4x^2} 
\frac{|V_{ts}V_{tb}|^2}{|V_{cb}|^2 f(z)} \, ,
\label{eq:bsnn}
\end{equation}
where $f(z)$ is given in Eq.~(\ref{phase}) and  
$z=m_c/m_b$ \cite{buras, ahmad}.
In the above result we summed over all neutrino flavors, therefore, the
$\sin^2\beta$ dependence cancels.
Using the experimental data
${\rm{Br}}(b \rightarrow c \mu^- {\overline {\nu_\mu}})=(10.5 \pm 0.5)\%$ 
\cite{cleo1},
we conclude 
\begin{equation}
{\rm{Br}}(b\rightarrow s \nu\overline{\nu})\lsim 
\frac{9.1\times 10^{-2}}{x^2}\, 
\end{equation}
after spanning the allowed values of the CKM matrix elements. 
It is interesting to notice that the predicted branching ratio is  
independent of $\sin^2\beta$ and $\sin^2\phi$ similar 
to the case of ${\rm{Br}}(K^+\rightarrow \pi^+ \nu\overline{\nu})$.
To agree with the experimental upper limit, 
which is $3.9\times 10^{-4}$ \cite{nardi}, 
it requires $x>15$, independent of the
parameters $\sin^2\phi$ and $\sin^2\beta$. 
Currently, for 
${\rm{Br}}(b\rightarrow s \nu\overline{\nu})$
to reach the same sensitivity as $\tau$ decay for $\sin^2\beta=0.5$, 
the measurement of ${\rm{Br}}(b\rightarrow s \nu\overline{\nu})$
has to be improved by a factor of 10.

Another interesting process to consider is the decay
$B_{s,d}\rightarrow \ell^+ \ell^-$. 
At tree level, the decay rate is given by
\begin{equation}
\Gamma(B_q\rightarrow \tau^+ \tau^- )=
\frac{G_F^2 f_{B_q}^2 m_{B_q} m_\tau^2 |V_{tb}V_{tq}|^2}{4\pi x^2} 
      {\left(\cos^2\beta -4\sin^2\theta\sin^2\phi\right)}^2
       {\left(1-\frac{4m_\tau^2}{m^2_{B_q}}\right)}^{3/2}\, .
\end{equation}
Using the values $\cos^2\beta=0.5$, $\sin^2\phi=0.04$, $m_{B_s}=5.369$ GeV, 
and $f_{B_s}=0.23$ GeV \cite{ali2}, 
the branching ratio  ${\rm{Br}}(B_{s}\rightarrow \tau^+ \tau^-)$ is given as 
\begin{equation}
{\rm{Br}}(B_{s}\rightarrow \tau^+ \tau^-)=\frac{4.6\times 10^{-3}}{x^2}\, .
\end{equation}
For $x\geq 48$, it corresponds to 
${\rm{Br}}(B_{s}\rightarrow \tau^+ \tau^-) \lsim 2.0\times 10^{-6}\, ,
$
which is of the same order as the SM prediction \cite{buras}. 
For the process $B_d\rightarrow \tau^+ \tau^-$, with $m_{B_d}=5.279$ GeV and
$f_{B_d}=0.18$ GeV \cite{ali2}, 
 the branching ratio ${\rm{Br}}(B_{d}\rightarrow \tau^+ \tau^- )$ is given as 
\begin{equation}
{\rm{Br}}(B_{d}\rightarrow \tau^+ \tau^- )=\frac{2.4\times 10^{-4}}{x^2}\, .
\end{equation}
For $x\geq 48$, it corresponds to 
${\rm{Br}}(B_{d}\rightarrow \tau^+ \tau^- ) \lsim 1.0\times 10^{-7}\, ,
$
which is again of the same order as the SM prediction \cite{buras}.

Next, consider the decay rates of 
$B_{s,d}\rightarrow \mu^+ \mu^-$. 
At tree level, the decay rate is given by
\begin{equation}
\Gamma(B_{q}\rightarrow \mu^+ \mu^- )= \frac{G_F^2 f_{B_{q}}^2 m_{B_q}
  m_\mu^2 |V_{tb}V_{tq}|^2 }{4\pi x^2}
        {\left(\sin^2\beta -4\sin^2\theta\sin^2\phi\right)}^2
        {\left(1-\frac{4m_\mu^2}{m^2_{B_q}}\right)}^{3/2}\, .
\end{equation}
Using the values $\sin^2\beta=0.5$, $\sin^2\phi=0.04$, and 
$f_{B_s}=0.23$ GeV \cite{ali2},
the branching ratio ${\rm{Br}}(B_{s}\rightarrow \mu^+ \mu^-)$ is given as
\begin{equation}
{\rm{Br}}(B_{s}\rightarrow \mu^+ \mu^-)=\frac{3.8\times 10^{-5}}{x^2}\, .
\end{equation}
For $x\geq 48$, we find
${\rm{Br}}(B_{s}\rightarrow \mu^+ \mu^-) \lsim 1.7\times 10^{-8}\, .
$
This result is smaller than the current experimental upper limit,
$2.6\times 10^{-6}\,$ \cite{abe}, by about two orders of magnitude.
Similarly, the branching ratio of ${B_d}\rightarrow \mu^+ \mu^-$ is 
given as 
\begin{equation}
{\rm{Br}}(B_{d}\rightarrow \mu^+ \mu^-)=\frac{2.0\times 10^{-6}}{x^2}\, .
\end{equation}
For $x\geq 48$, we find
${\rm{Br}}(B_{d}\rightarrow \mu^+ \mu^-) \lsim 8.8\times 10^{-10}\, .
$
Again, this result is smaller than the experimental 
upper limit, $8.6\times 10^{-7}\,$  \cite{abe},
by three orders of magnitude.

Finally, we note that in this model, with lepton mixing, 
it is possible to have the decay modes of 
$B_{d,s}\rightarrow \mu^\pm \tau^\mp$, which are absent in the SM.
In the limit of ignoring the mass difference between $\tau$
and $\mu$, with maximal lepton mixing, their branching ratios are 
about twice of those for the $\tau \tau$ modes. 
Hence, detecting such non-standard decay modes can further 
constrain the model, especially on the lepton mixing parameter
$\sin^2 \beta$.

In conclusion, under the down-type quark mixing scenario, the 
decay width of $K^+ \rightarrow \pi^0 e^+ \nu_e$ is not modified
at tree level, if assuming the third family lepton does not mix with
the first family lepton.
The branching ratios of $K^+ \rightarrow \pi^+ \nu\overline{\nu}$ 
can be an order of magnitude larger than the SM prediction,
and can be tested at Kaon factories. 
The effect to the $K^0$-$\overline{K^0}$ mixing is of the same order as the 
SM prediction, which can prove to be useful if the long-distance contribution
can be better understood theoretically.
Similarly, the branching ratio  
${\rm{Br}}(b\rightarrow s \nu\overline{\nu})$ is modified and can be an
order of magnitude larger than the SM prediction.
Furthermore, since the branching ratios  
${\rm{Br}}(K^+ \rightarrow \pi^+ \nu\overline{\nu})$ and
${\rm{Br}}(b\rightarrow s \nu\overline{\nu})$
do not depend on $\sin^2\phi$ and $\sin^2\beta$, 
they can be extremely useful in constraining the remaining parameter $x$.

The current data on the branching ratios of 
$B_{s,d} \rightarrow \tau^- \tau^+, \mu^- \mu^+$ and
$b \rightarrow s \mu^- \mu^+, s e^- e^+$ 
does not impose a better constraint on the model 
than that by the $Z$-pole measurements. 
However, with a much larger statistics of
the data in the $B$-factories, we expect it to be improved.
Since this model also predicts non-SM decay modes, such as
$b \rightarrow s \mu^\pm \tau^\mp$ and 
$B_{s,d} \rightarrow \mu^\pm \tau^\mp$,
with comparable branching ratios,
they should be measured to test the model prediction 
on the lepton mixing dynamics (i.e., $\sin^2\beta$ dependence).
For the range of the parameter $x$ consistent with the $Z$-pole
data, it is found that in this model
a new contribution to the $B^0$-$\overline{B^0}$ mixing can reach the 
range of $150\%$--$360\%$. 

As a summary, in Table 3 we tabulate the predictions of this model for 
various decay processes. Two cases are considered, one for 
$\sin^2\beta=0.0$ and $x=20$, another for $\sin^2\beta=0.5$ and $x=48$. 
For both cases we set $\sin^2\phi=0.04$.

\subsection{Mixing in the Up-Quark Sector}

\indent

 In this section we assume that no mixing occurs in
the down-type quarks, so that $L_d^{\dagger }GL_d=G$, and
$L_u^{\dagger}GL_u=VGV^{\dagger}$.
In this case, the quark interactions to the gauge bosons are given 
in Eq.~(\ref{qu2}) with the above substitutions. 
Similar to the case of down-type quark mixing,
the FCNC interactions are completely determined by 
the CKM matrix $V$ and the two parameters $\sin^2\phi $ and $x$. 
The matrix $VGV^{\dagger}$
can be explicitly written as
\begin{equation}
VGV^\dagger=\left( 
\begin{array}{ccc}
|V_{ub}|^2 & V_{ub}V_{cb}^{*} & V_{ub}V_{tb}^{*} \\ 
V_{cb}V_{ub}^{*} & |V_{cb}|^2 & V_{cb}V_{tb}^{*} \\ 
V_{tb}V_{ub}^{*} & V_{tb}V_{cb}^{*} & |V_{tb}|^2
\end{array}
\right) .
\end{equation}
Again, because the elements of the 
matrix $V^{\dagger }GV$
are naturally small, we do not expect large
effects in the FCNC processes. 

\subsubsection{Charged-Current Phenomenology}

\indent

The non-standard contribution to charged-current processes is given
by Eq.~(\ref{eq:cc}). 
In terms of the mass
eigenstates and assuming no mixing in the down-type quarks, 
$j_h^{-}$ contains the
following charged currents: 
\begin{equation}
j_h^{-}=V_{ub}\overline{\ u_L}\gamma _\mu \ b_L\ ,\
        V_{cb}\ \overline{c_L}\gamma _\mu \ b_L\ ,\ 
        V_{tb}\ \overline{t_L}\gamma _\mu \ b_L.
\end{equation}
It is important to note that only the 
$b$ quark, among the down-type quarks, appears in $j_h^{-}$, which implies
that new effects in the charged currents must involve the $b$ quark. 
Because the non-standard contribution has the form $j_h^{+}j_h^{-}$, 
the only non-vanishing effect we expect in the 
pure hadronic charged-current interaction 
(at the leading order in $1/x$)
is that with a $b$ quark in both currents, i.e., with  $\Delta B=0$.
Therefore, no new effect is expected in any of the hadronic 
decay channel of $K$, $D$, and $B$ mesons.

Next, let us consider the semi-leptonic decay processes. 
The relevant hadronic currents are 
\begin{equation}
V_{ub}\ \overline{u_L}\gamma _\mu \ b_L\ ,\ V_{cb}\ \overline{c_L}
\gamma _\mu \ b_L,
\end{equation}
while the relevant leptonic currents are 
\begin{equation}
\sin^2\beta \ \overline{\mu _L}\ \gamma _\mu \ {\nu _\mu }_L\ ,\
        \cos^2\beta \ \overline{\tau _L}\ \gamma _\mu\ {\nu _\tau}_L\ ,\ 
\sin \beta \cos \beta \ \overline{\mu _L}\ \gamma _\mu \ {\nu_\tau }_L\ ,\ 
\sin \beta \cos \beta \ \overline{\tau _L}\ \gamma _\mu \ {\nu }_{\mu L}\ .
\end{equation}
It is clear that
new effects in the charged-current semi-leptonic decays 
are only expected in the
$b$-quark system. Explicitly, the decay processes 
$b\rightarrow u\ (\mu ,\tau )(\nu _\mu ,\nu _\tau )$ 
will receive new contributions induced by the following interaction terms:
\begin{eqnarray*}
&&\ \frac{2\sqrt{2}G_F\ V_{ub}\ }x\left\{ \sin ^2\beta 
\left( \ \overline{u_L}\gamma _\mu \ b_L\right) 
\left( \overline{\mu _L}\ \gamma _\mu \ {\nu _\mu }
_L\right) ,\ \cos ^2\beta \left( \ \overline{u_L}\gamma _\mu \ b_L\right)
\left( \overline{\tau _L}\ \gamma _\mu \ {\nu _\tau }_L\right) \right\},  \\
&&\ \frac{2\sqrt{2}G_F\ V_{ub}\ \sin \beta \cos \beta }x\left\{ \left( \ 
\overline{u_L}\gamma _\mu \ b_L\right) \left( \overline{\mu _L}\ \gamma _\mu
\ {\nu _\tau }_L\right) ,\ \left( \ \overline{u_L}\gamma _\mu \ b_L\right)
\left( \overline{\tau _L}\ \gamma _\mu \ {\nu _\mu}_{L}\right)
\right\} .
\end{eqnarray*}
A similar expression for the $b$ decay to charm can 
be obtained with $V_{ub}$ replaced by $V_{cb}$. 
Hence, we expect an increase in the $b$-quark
semi-leptonic decays as compared to the SM. As an example, 
the branching ratio of
$B_d^0\longrightarrow D^{-}\ell^{+}\nu $ and  
$B_s^0\longrightarrow D_s^{-}\ell^{+}\nu $ is predicted to be
\begin{equation}
{\rm{Br}(B^0\longrightarrow D^{-}\ell ^{+}\nu)}=
{\rm{Br}}^{\rm{SM}}(B^0\longrightarrow D^{-}\ell ^{+}\nu)
\left( 1+\frac{2}{x}\right) ,
\end{equation}
where all the three lepton (including neutrino) flavors are included.
(Note that the $\sin^2\beta$ dependence cancels.)
With $x>20$, imposed by $Z$-pole data, we do not expect the 
new physics effect to exceed $10\%$.
Because of the large uncertainty (exceeding 
$25\%$ \cite{data}) of present data,
these processes do not offer a stringent constraint on the model.
With more statistics of the future data, these decay processes can be useful
for constraining the parameter $x$.

Under this scenario, we conclude that the values of the CKM 
matrix elements extracted from tree-level processes not involving 
the $b$ quark 
are not modified by the model. 
In other words, the extracted values of the CKM elements, $V_{ud}$, $V_{us}$,
$V_{cd}$, and $V_{cs}$, for the SM and this model coincide. 
However,
the matrix elements $V_{ub}$ and $V_{cb}$ are modified slightly. To explore
this effect, let us consider the transition 
$b\rightarrow u \mu^- {\overline {\nu_\mu}}$. 
Its amplitude is modified with $V_{ub}$ replaced by 
 $V_{ub}(1+{\sin^2 \beta}/{x})$. Therefore, the extracted 
experimental value of $V_{ub}$, assuming the validity of the SM, is  
equivalent to the quantity $V_{ub}(1+{\sin^2\beta}/{x})$ in this model. 
{}From the data, the unitarity condition for the SM reads as \cite{ali1} 
\begin{equation}
{|V_{ud}|}^2 +{|V_{us}|}^2 +{|V_{ub}|}^2=0.997 \pm 0.002\, .
\end{equation}
Hence, at the $2\sigma$ level, we conclude that $x \geq  0.05\sin^2\beta$.
It is clear that the unitarity condition does not add any useful 
constraint on the model after testing against the $Z$-pole 
data which requires $x>20$.

\subsubsection{Neutral-Current Phenomenology}

\indent

First, let us consider neutral-current processes 
of hadron-hadron interactions. In this case, the relevant neutral currents are 
\begin{equation}
j_h^3=\overline{t_L}\gamma _\mu \left(\frac{1}{2}\right) t_L\ ,
\overline{b_L} \gamma _\mu \left(-\frac{1}{2}\right) b_L,
\end{equation}
written in terms of the weak eigenstates, which yields
the following four-fermion interaction current
in terms of the mass eigenstates:
\begin{equation}
|V_{ub}|^2\ \overline{u}_L\gamma _\mu \ u_L \, ,
|V_{cb}|^2\ \overline{c}_L\gamma _\mu \ c_L\, ,
V_{ub}V^*_{cb}\ \ \overline{u_L}\gamma_\mu  c_L\, . \label{last}
\end{equation}
In the four-fermion neutral-current interaction we notice that the $d$ 
and $s$ quarks will appear only through the electromagnetic current 
$J_{\rm{em}}$ (cf. Eq.~(\ref{nc})). 
Because of the structure of the neutral-current
interaction, new physics can only contribute to  
processes with $\Delta B=0$. 
Thus, neither the $B$ hadronic decay nor the $B^0$-$\overline{B^0}$ mixing
is modified at tree level in this model. Similarly, we conclude that new
effects must have $\Delta S=0$ in pure hadronic interaction.
Therefore, $K^0$-$\overline{K^0}$ mixing is not modified. 
We conclude that new physics effects, in the pure hadronic 
decay modes, are only expected in the $c$-quark decay channels. 
Nevertheless, these new physics effects are naturally small because
the FCNC couplings predicted by this model at tree level are suppressed by 
products of CKM matrix elements. 

Second, let us consider the semi-leptonic decays. Again, 
we do not expect any new
effects in the $b$-quark semi-leptonic decays because of the requirement 
$\Delta B=0$. Effects are only expected in the charm decay where we get
interactions of the form
\begin{equation}
\frac{\sqrt{2}G_F}xV_{ub}^{*}V_{cb}\ \ (\overline{u_L}\gamma ^\mu c_L)\left(
\sin ^2\beta (\overline{\mu }_L\gamma _\mu \ \mu _L)-2\sin ^2\phi 
\sin^2\theta \ (\overline{\mu }\ \gamma _\mu \ \mu )\right) .
\end{equation}
Because of the large suppression factor $V_{ub}^{*}V_{cb}$
and large error on present experimental data,
it is extremely difficult to gain any further information about the model
from the semi-leptonic decay channels of charm hadrons.

The only suspected new effect in the $b$-quark system is through the 
$\Upsilon(1S)$ decay. In this case, the decay proceeds through 
$b\overline{b}\rightarrow \gamma, Z, Z^\prime \rightarrow \mu ^{+}\mu ^{-}$. 
At tree level, the new contribution is expressed through the
interaction term
\begin{equation}
\frac{\sqrt{2}G_F}{x} \left( (\overline{b_L}\gamma^\mu b_L)-
\frac {2}{3} \sin^2\phi \sin^2\theta 
(\overline{b}\ \gamma _\mu \ b)\right) 
\left( \sin^2\beta (\overline{\mu }_L\gamma _\mu \mu _L)-
2\sin ^2\phi \sin ^2\theta (\overline{\mu }\ \gamma _\mu \ \mu )\right) . 
\end{equation}
For very small values of $\sin^2\phi $ and for large possible lepton
mixing (i.e., large $\sin^2\beta$), we
can approximate the above interaction 
relevant to $\Upsilon(1S)\rightarrow \mu^+\mu^-$ as
\begin{equation}
\frac{\sqrt{2}G_F \sin^2\beta}{x} \left( \overline{b_L}\gamma^\mu b_L\right) 
\left(\overline{\mu }_L\gamma _\mu \ \mu _L\right) . 
\end{equation}
Needless to say, the
dominant contribution to the $\Upsilon(1S)$ decay width is coming from
the photon exchange. The non-standard contribution predicted by this
model can be estimated as follows.
The amplitude $\Upsilon(1S)\rightarrow \ell^+\ell^-$ can in general be 
written as \cite{hill2}
\begin{equation}
T(\Upsilon(1S)\rightarrow \ell^+\ell^-)=-\frac{4\pi \alpha}{3M_\Upsilon^2}
<0|(bb)_V|\Upsilon>\left[r_V(\ell\ell)_V + r_A(\ell\ell)_A\right]\, .
\end{equation}
For the dominant photon contribution, 
$r_V=1$ and $r_A=0$. In the case of $\tau$ 
lepton mode, these couplings in the proposed model 
will be modified into
\begin{equation}
r_V=1-\frac{3M_\Upsilon^2 \cos^2\beta}{16\sin^2\theta M_W^2 x}\, ,
\end{equation}
\begin{equation}
r_A= \frac{3M_\Upsilon^2 \cos^2\beta}{16\sin^2\theta M_W^2 x}\, .
\end{equation}
A similar relation holds for the $\mu$ lepton mode, but with $\cos^2\beta$
replaced by $\sin^2\beta$.
The ratio of the $\tau$ lepton decay rate to the $\mu$ lepton decay rate 
is  
\begin{equation}
\frac{\Gamma(\Upsilon(1S)\rightarrow \tau^+ \tau^-)}
     {\Gamma(\Upsilon(1S)\rightarrow \mu^+ \mu^-)} =
\sqrt{1-4\frac{m_\tau^2}{M_\Upsilon^2}} 
\left(1+2\frac{m_\tau^2}{M_\Upsilon^2}\right)
\left[1 -\frac{3M_\Upsilon^2}{8\sin^2\theta M_W^2 x} 
\left(\cos^2\beta-\sin^2\beta\right) \right]\, ,
\end{equation}
which amounts to new effect of the order   
\begin{equation}
  -2.3 \times 10^{-2}\frac{1}{x}\left(\cos^2\beta-\sin^2\beta\right) \, .
\end{equation}
Therefore, the expected maximal deviation is less than $\pm0.1\%$, 
for $x\geq 20$. (It vanishes for the maximal lepton mixing scenario,
i.e. for $\sin^2\beta=0.5$.)
The current experimental error is at the percent level \cite{data},
so that it does not provide additional constraints on the model.
However, it is interesting to notice that the sign of the deviation 
is governed by the difference $\cos^2\beta -\sin^2\beta$. Future measurements
with a much less error can be used to determine the 
lepton mixing angle. 
Finally, we note that the non-SM decay mode
$\Upsilon(1S)\rightarrow \mu^\pm \tau^\mp$ is expected by this model with
a branching ratio less than $4\times 10^{-10}$.
Since this decay process is not allowed by the SM, 
it can provide a significant constraint on the 
lepton mixing parameter $\sin^2\beta$.

The above discussion is valid for tree-level contributions. 
We now consider whether one-loop effects can be significant to some
observables, such as the $K^0$-$\overline{K^0}$, 
$B^0$-$\overline{B^0}$ mixing, and the decay branching ratio of
$b \to s \gamma$.

\subsubsection{One-Loop Effects}

\indent

In the SM, $K^0$-$\overline{K^0}$ and $B^0$-$\overline{B^0}$ mixing are 
induced via one-loop $W$-$W$ exchange box diagrams. 
In this model and under the scenario of a trivial $L_d$, 
the $K^0$-$\overline{K^0}$ and $B^0$-$\overline{B^0}$  
mixing can occur at the one-loop level
through box diagrams involving the exchange of $W$ and/or $W^\prime$ 
gauge bosons. In addition 
to the SM diagrams, there are four box 
diagrams with one $W$ and one
$W^\prime$ exchange. Diagrams with two $W^\prime$ exchange do not contribute
at the order $1/x$ but at the higher order $1/x^2$.   

For the case of $K^0$-$\overline{K^0}$ mixing, 
we calculate the one-loop amplitude and compare with the short distance 
contribution of the SM. Substituting the values
of $M_W$, $m_t$, $m_c$, and the $V_{CKM}$ elements \cite{data}, we find 
\begin{equation}
\frac{T}{T^{\rm{SM}}}=\frac{\Delta M}{{\Delta M}^{\rm{SM}}}
\lsim 3.4\times 10^{-3} \frac{\sin ^4\phi }x .
\end{equation}
Since constraints imposed by the $Z$-pole data require 
$\sin^2\phi/x < 0.3\%$, new effect to the $K^0$-$\overline{K^0}$
are extremely small and of no relevance to the discussion. 

Next, let us consider the $B_q$-$\overline{B_q}$ mixing. 
The leading SM one loop amplitude is given in Eq.~(\ref{B_mix}). 
Similar to the $K^0$-$\overline{K^0}$ case, we include 
the additional box diagrams that contribute at the 
order $1/x$, we find
\begin{equation}
\frac{T}{T^{\rm{SM}}}=\frac{\Delta M_q}{{\Delta M_q}^{\rm{SM}}}=
\frac{2\sin^2\phi}{x} -2.24\frac{\sin^4\phi}{x}\, .
\end{equation}
Using the $Z$-pole constraint ($x>20$ for $\sin^2\phi >0.04$), 
we expect new effect to the amplitude not to exceed $0.4\%$
relative to the SM. 
Therefore, we do not expect large new effect to the 
$B_q$-$\overline{B_q}$ mixing at 
the one-loop level for the case of trivial $L_d$.

Finally, let us consider the decay of $b\rightarrow s \gamma$.
The SM amplitude for the process $b \rightarrow s\, \gamma$ 
is given by \cite{fuji}
\begin{equation}
T^{\rm{SM}}=\frac 1{16\pi ^2}\frac{m_b}{M^2} \left( \frac{e g^2}{4}
V_{ts}^{*}V_{tb}\right) \left[ \overline{u_s}(1+\gamma _5)\left(
2p.\varepsilon -\varepsilon _\mu \gamma ^\mu \right) u_b\right]
\left\{T_1+T_2\right\} , 
\end{equation}
where
\begin{equation}
T_1=\frac 1{(y-1)^4}\left[ \frac{y^4}2+\frac 34y^3-\frac 32y^2+\frac
14y-\frac 32y^3\ln y\right] , 
\end{equation}
\begin{equation}
T_2=\frac{Q_t}{(y-1)^4}\left[ \frac{y^4}4-\frac 32y^3+\frac 34y^2+\frac
12y+\frac 32y^2\ln y\right] , 
\end{equation}
and $Q_t=2/3$ is the electric charge of the top quark.

In this model, the $b \rightarrow s \gamma$ 
amplitude will be slightly changed due to the modified
couplings. The only diagrams we need to consider are the usual $W$ exchange 
penguin diagrams. Since the fermion couplings are slightly modified, 
these diagrams
will contain an extra contribution with respect to the SM. 
The penguin diagrams
with $W^\prime$ exchange do not contribute to the order $1/x$.
We calculate the new amplitude as predicted by the model and compare 
it to the SM one. After substituting the values of $M_W$ and $m_t$, we find 
\begin{equation}
\frac{T}{T^{{\rm SM}}}= -1.7\frac{\sin^2\phi}{x}+
                         1.4\frac{\sin^4\phi}{x}\, .
\end{equation}
Therefore, for the $Z$-pole limit ($x>20$), we expect new contribution
not to exceed $0.3\%$ of the SM.

In conclusion, under the up-type quark mixing scenario, this model
does not modify the $K^0$-$\overline{K^0}$, 
$B^0$-$\overline{B^0}$ mixing, 
and the decay width of $b \rightarrow s \gamma$ at tree level.
The one-loop effects to these observables are small, and 
do not exceed the level of $0.4\%$ of the SM values. 
In general, we conclude that the up-type quarks mixing 
scenario can hardly
be examined against the low-energy data available so far. 

\subsection{The general mixing scenario}

\indent

In this section, we consider the general case of both types of quark 
mixing, i.e,  
where both $L_u$ and $L_d$ are  non-trivial.
The charged-current mixing matrix $V$ is defined the same as 
before, $V=L_u^{\dagger} L_d$. 
The interaction Lagrangian can be expressed using  
two matrix structures, such as $V$ and $L_d^{\dagger}G L_d$. 
In this case $L_u^\dagger G L_u=VL_d^{\dagger}G L_d V^\dagger$. 
Therefore, under the general mixing scenario,
there are additional free 
parameters in comparison with the previously discussed two cases. 
The additional 
parameters  appear in the matrix $L_d^\dagger GL_d$,
where
\begin{equation} 
L_d^\dagger GL_d= {\pmatrix{|d_{31}|^2 & d_{31}^\ast d_{32} & d_{31}^\ast 
d_{33} \cr 
              d_{31}d_{32}^\ast & |d_{32}|^2 & d_{32}^\ast d_{33} \cr 
              d_{31}d_{33}^\ast &  d_{32}d_{33}^\ast & |d_{33}|^2 \cr}}\,.
\end{equation}
Since unitarity condition 
implies that $|d_{31}|^2 + |d_{32}|^2 +|d_{33}|^2 =1$, 
there are only two additional free parameters which will be assumed to be real
numbers hereafter. (Additional phases can be generated which would  
signal a new source of CP violation.) 

The general case is more tolerant to accommodate low-energy data because 
of the additional parameters. 
Nevertheless, as to be shown later, we can set significant constraints on 
some combination of those additional parameters. 
In the following, we shall examine a few relevant tree-level
FCNC processes.

As a start, we consider the decay
$K^+ \rightarrow \pi^+ \nu {\overline \nu} $.
As discussed before, this process can occur in this model at tree level 
through the flavor-changing neutral current
$s\rightarrow d\, Z\rightarrow d \nu \overline{\nu}$. 
The branching ratio of this process can be obtained from 
the ratio 
\begin{equation}
R=\frac{{\rm{Br}}(K^+ \rightarrow \pi^+ \nu\overline{\nu})}
       {{\rm{Br}}(K^+ \rightarrow \pi^0 e^+ \nu_e )}=
\frac{1}{4x^2} \left( \frac{|d_{31}|^2|d_{32}|^2} {|V_{us}|^2}\right)\, ,
\end{equation}
which noticeably is independent of the parameters
$\sin^2\beta$ and $\sin^2\phi$. Therefore, the ratio $R$ can be
used to directly set a limit 
on $|d_{31}d_{32}|/x$, without any assumptions regarding other 
parameters.
If we compare this result with the published result of the E787 
collaboration \cite{e787}, 
${\rm{Br}}(K^+ \rightarrow \pi^+ \nu\overline{\nu})= 4.2^{+9.7}_{-3.5}
\times 10^{-10}$,
we obtain the $2\sigma$ level constraint: 
\begin{equation}
\frac{|d_{31} d_{32}|}{x} \lsim 10^{-4}\, .
\label{eq:kpnn}
\end{equation}
For $x=20$, the smallest value of $x$ consistent with the $Z$-pole data,
it requires $|d_{31}d_{32}|< 2 \times 10^{-3}$.

Now we consider the new effect to the $K^0$-$\overline{K^0}$ mixing. 
A straightforward calculation
of the tree-level amplitude as compared with the SM short 
distance contribution gives
\begin{equation}
\frac{T}{T^{\rm{SM}}}=\frac{\Delta M}{{\Delta M}^{\rm{SM}}}
\approx 1\times 10^{7} 
\frac{{\rm{Re}}{\left( d^*_{31} d_{32}\right)}^2}{x}\, .
\end{equation}
Combined with the previous constraint derived from 
$K^+ \rightarrow \pi^+ \nu {\overline{\nu}} $, we conclude 
\begin{equation}
\frac{\Delta M}{{\Delta M}^{\rm{SM}}} \lsim 1 \times 10^{3}
      |d_{31} d_{32}| \, .
\end{equation}
Hence, the combination of the 
$K^+ \rightarrow \pi^+ \nu {\overline{\nu}} $ and
$K^0$-$\overline{K^0}$ mixing data directly constrains 
the magnitude of $|d_{31} d_{32}|$ because the explicit $x$ dependence cancels.
If $x=20$, the non-standard contribution in 
$K^0$-$\overline{K^0}$ mixing can be as large as twice  
the SM short distance contribution.

Next, we use bottom physics data to constrain 
the second additional free parameter. 
Consider the decay rate $b\rightarrow s \nu \overline{\nu}$. 
The expected branching ratio,
which is independent of the parameters $\sin^2\phi$ and 
$\sin^2\beta$, is given by
\begin{equation}
\frac{{\rm{Br}}(b\rightarrow s \nu \overline{\nu})}
     {{\rm{Br}}(b \rightarrow c \mu^- \overline{\nu_\mu})}
=\frac{1}{4x^2} 
\frac{|d_{32} d_{33}|^2}{|V_{cb}|^2 f(z)} \, ,
\end{equation}
where $f(z)$ is given in Eq.~(\ref{phase}) and $z=m_c/m_b$.
Using the experimental data,
${\rm{Br}}(b \rightarrow c \mu^- {\overline {\nu_\mu}})=(10.5 \pm 0.5)\%$
\cite{cleo1}
and
${\rm{Br}}(b\rightarrow s \nu\overline{\nu})
< 3.9\times 10^{-4}$ \cite{nardi}, 
we obtain the constraint 
\begin{equation}
\frac{|d_{32} d_{33}|}{x} < 2.9 \times 10^{-3} \, ,
\end{equation}
For $x = 20$, it requires $|d_{32} d_{33}| < 0.06$.
Next, we consider the $B^0_s$-$\overline{B_s^0}$ mixing. 
A straightforward calculation
of the new physics effect to the $B^0_s$-$\overline{B_s^0}$ mixing 
compared with the SM contribution gives
\begin{equation}
\frac{T}{T^{\rm{SM}}}=
\frac{\Delta M_{B_s}}{{(\Delta M_{B_s})}^{\rm{SM}}}
\approx 3.4\times 10^{4} 
\frac{{\rm{Re}}{\left( d^*_{32} d_{33}\right)}^2}{x} \, .
 \end{equation}
When combined with the previous constraint derived from the 
measurement of ${\rm{Br}}(b\rightarrow s \nu\overline{\nu})$,
it yields
\begin{equation}
\frac{\Delta M_{B_s}}{{(\Delta M_{B_s})}^{\rm{SM}}}
\lsim 100 |d_{32} d_{33}|  \, .
 \end{equation}
Hence, the combination of the 
$b\rightarrow s \nu\overline{\nu}$ and 
$B^0_s$-$\overline{B_s^0}$ mixing data directly constrains 
the magnitude of $|d_{32} d_{33}|$ because the explicit $x$ dependence cancels.
For $x=20$, the non-standard contribution to 
$B^0_s$-$\overline{B_s^0}$ mixing can be as large as six times 
the SM short distance contribution.

Given the constraints on $|d_{31} d_{32}|$, $|d_{32} d_{33}|$, and 
 the unitarity condition on the matrix $L_d$, 
one can derive the allowed space of the parameters  
$|d_{31}|$, $|d_{32}|$, and $|d_{33}|$.
It is interesting to notice that in the SM
neither $L_u$ nor $L_d$ can be separately determined, and only
the CKM matrix $V$, which is the product of 
$L_u^\dagger$ and $L_d$, can be measured experimentally.
However, in this model, the elements in the third column of 
the $L_{u,d}$ mixing matrices can be determined, and can be 
further constrained by including other low-energy data.
Unfortunately, in general, those observables depend also on some other
parameters, such as $\sin^2\phi$ and $\sin^2\beta$, of the model.
Some of them are discussed below.

The expected branching ratio for $b\rightarrow s \mu^+\mu^-$ 
is given by
\begin{equation}
\frac{{\rm{Br}}(b\rightarrow s \mu^-\mu^+)}
{{\rm{Br}}(b \rightarrow c \mu^- {\overline{\nu_\mu}})}=
\frac{1}{4x^2} 
\frac{|d_{32}|^2 |d_{33}|^2}{|V_{cb}|^2 f(z)} 
\left( \sin^4\beta-4\sin^2\beta\sin^2\phi\sin^2\theta +
        8\sin^4\theta\sin^4\phi \right)\, .
\end{equation}
Using the CLEO data \cite{cleo1}, 
we obtain 
\begin{equation}
\frac{|d_{32} d_{33}|}{x} < 2.4 \times 10^{-3} \, ,
\end{equation}
for $\sin^2\beta=0.5$ and $\sin^2\phi=0.04$.
For $x = 48$, the minimal value of $x$ consistent with $Z$-pole data and 
$\tau$ life-time, it requires $|d_{32} d_{33}| < 0.12$.

Next, consider the decay rate of
$B_{s,d}\rightarrow \mu^+ \mu^-$. 
The tree level contribution gives
\begin{equation}
\Gamma(B_{s}\rightarrow \mu^+ \mu^- )= \frac{G_F^2 f_{B_{q}}^2 m_{B_q}
  m_\mu^2 |d_{32} d_{33}|^2 }{4\pi x^2}
        {\left(\sin^2\beta -4\sin^2\theta\sin^2\phi\right)}^2
        {\left(1-\frac{4m_\mu^2}{m^2_{B_q}}\right)}^{3/2}\, .
\end{equation}
For $\sin^2\beta=0.5$ and $\sin^2\phi=0.04$, 
the branching ratio ${\rm{Br}}(B_{s}\rightarrow \mu^+ \mu^-)$ is 
\begin{equation}
{\rm{Br}}(B_{s}\rightarrow \mu^+ \mu^-)=0.018 \frac{|d_{32}d_{33}|^2}{x^2}\, .
\end{equation}
Comparing this result with the experimental upper limit \cite{abe},
we obtain 
\begin{equation}
\frac{|d_{32}d_{33}|}{x} < 1.2\times 10^{-2}\, .
\end{equation}
This constraint is not as strong as the one obtained from 
$b \rightarrow s \mu^+\mu^-$,
the latter is stronger by one order of magnitude.

The branching ratio of $B_{d}\rightarrow \mu^+ \mu^-$, for
$\sin^2\beta=0.5$ and $\sin^2\phi=0.04$, is 
\begin{equation}
{\rm{Br}}(B_{d}\rightarrow \mu^+ \mu^-)=0.01 \frac{|d_{31}d_{33}|^2}{x^2}\, .
\end{equation}
Comparing this result with the experimental upper limit \cite{abe},
we obtain 
\begin{equation}
\frac{|d_{31}d_{33}|}{x} < 9.1\times 10^{-3}\, .
\end{equation}
For $x=48$, it yields $|d_{31}d_{33}| < 0.44$.

The new physics effect to the $B^0_d$-$\overline{B_d^0}$ mixing 
compared with the SM contribution can be written as 
\begin{equation}
\frac{\Delta M_{B_d}}{({\Delta M_{B_d})}^{\rm{SM}}}=
\frac{T}{T^{SM}}
\approx 3.6\times 10^{5} 
\frac{{\rm{Re}}{\left( d^*_{31} d_{33}\right)}^2}{x}\, .
 \end{equation}
When combined with the above constraint derived from the 
decay $B_d \rightarrow \mu^+\mu^-$ \cite{abe},
it yields
\begin{equation}
 \frac{\Delta M_{B_d}}{{(\Delta M_{B_d})}^{\rm{SM}}}
 \lsim 3.3\times 10^{3} |d_{31} d_{33}|  \, .
 \end{equation}
If we consider the values $x=48$, $\sin^2\beta=0.5$, and $\sin^2\phi=0.04$, 
then $|d_{31}d_{33}| < 0.44$ and  
${\Delta M_{B_d}}/{{(\Delta M_{B_d})}^{\rm{SM}}} \lsim 1450$,
which implies that the current measurement of 
${\rm{Br}}(B_{d}\rightarrow \mu^+ \mu^-)$ is 
not useful to constrain this model, 
and new physics effect to $B^0_d$-$\overline{B_d^0}$ mixing
can be much larger than the SM prediction.
A precision measurement of $B^0_d$-$\overline{B_d^0}$ mixing 
will be extremely valuable to test this model with the scenario
that both the up- and down-type quarks can mix in their mass
eigenstates.

Similar to the discussions given for the other two scenarios, this model
also allows lepton number violation processes, such as 
$B_{s,d}\rightarrow \mu^\pm \tau^\mp$ and 
$b \rightarrow s \mu^\pm \tau^\mp$.
Since their branching ratios are of the same order as those for the 
$\tau^+\tau^-$ mode, they can be very useful for further testing the model.
In conclusion, under the general mixing scenario,
the model requires two additional free (real) parameters, 
although additional phases can be introduced to 
generate a new source of CP violation.
Depending on the values of the parameters, 
sizable effects in various FCNC processes are expected.
In Table 4, we summarize the results of this 
section by giving the constraints on the mixing parameters as extracted from
different experiments. 

\section{Conclusions}
 
\indent

In this work, we revisit the model in Ref.~\cite{ehab5}, and update
the constraints on this model from the $Z$-pole data at LEP/SLC.
We find that the heavy gauge boson mass is 
bounded from below to be about $1.7$ TeV at the $2\sigma$ level.
The parameter $x$, the square of the ratio of the 
two VEVs involved in the breaking pattern of the gauge symmetry, 
is larger than 20 assuming no lepton mixing, and 48 with the
maximal possible lepton mixing between $\mu$ and $\tau$. 
Given that, we study the potential of the new physics effect predicted 
by this model to low-energy data with zero momentum transfer,
such as $K$ and $B$ physics.
We concentrate on the region where $x$ is large.
Using an effective current-current interaction Lagrangian, we
systematically examine the possible new physics effects in the
charged-current and the neutral-current interactions.
We show that FCNC couplings in this model can be written as the  
product of CKM matrix elements, so that FCNC processes are
naturally suppressed.
To examine how well low-energy data can further test this model,
we have separately studied three different scenarios of quark mixing.
 
Assuming the third family lepton does not mix with the 
first family lepton, the partial decay width of
$\mu \rightarrow eee$ and $\mu \rightarrow e \gamma$
will not be modified.  
The current data on the measurement of the
ratio $\Gamma(\tau \rightarrow \mu \overline{\nu _\mu }\nu _\tau)/
\Gamma(\tau \rightarrow e \overline{\nu _e }\nu _\tau)$ places the
strongest constraint on the parameter $x$, which is even better than
$Z$-pole constraint for $\sin^2\phi <0.1$ (cf. Figure 2). 
The lepton number violation
process $\tau \rightarrow \mu \mu \mu $ is also significant and gives 
a compatible constraint as the above measurement. 
On the other hand, given the current experimental data, 
the decay process $\tau\rightarrow \mu \gamma$ 
 and the measurement of the anomalous magnetic dipole moment of the muon 
are not yet significant in constraining the model.
If the above discussed processes can be measured to a better accuracy 
in future experiments, they will play a more significant role in
testing the model considered in this work.

Under the down-type quark mixing scenario, the 
decay width of $K^+ \rightarrow \pi^0 e^+ \nu_e$ is not modified
at tree level, if assuming the third family lepton does not mix with
the first family lepton.
The branching ratio of $K^+ \rightarrow \pi^+ \nu \overline{\nu}$ can be 
an order of magnitude larger than the SM prediction. In that case, 
it can be tested at Kaon factories. 
The effect to the $K^0$-$\overline{K^0}$ mixing is of the same order as the 
SM prediction, which can only be useful if the long distance contribution
can be better understood theoretically.
Furthermore, since the above observables do not depend on the parameter
$\sin^2\phi$, they can directly constrain the parameter $x$ of the model.
The current data on the branching ratios of 
$B_{s,d} \rightarrow \tau^- \tau^+, \mu^- \mu^+$ and
$b \rightarrow s \mu^- \mu^+, s e^- e^+,  s\nu\overline{\nu}$ 
do not impose a better constraint on the model 
than that by the $Z$-pole measurements. 
However, with a much larger statistics of
the data in $B$ (Beauty) factories, we expect it to be improved.
Since this model also predicts the non-SM decay modes, such as
$b \rightarrow s \mu^\pm \tau^\mp$ and 
$B_{s,d} \rightarrow \mu^\pm \tau^\mp$,
with comparable branching ratios,
they should be measured to test the model prediction 
on the lepton mixing dynamics (i.e., $\sin^2\beta$ dependence).
For the range of the parameter $x$ consistent with the $Z$-pole
data, it is found that in this model
a new contribution to the $B^0_q$-$\overline{B^0_q}$ mixing can reach the 
range of $150\%$--$360\%$. Hence, this measurement is useful 
for testing the model.
As a summary to  this scenario, in Table 2 we give the lower bound on 
the parameter $x$ derived from 
including the low energy data as well as the $Z$-pole data.
We consider
two cases. Case I: No lepton mixing ($\sin^2\beta=0$). Case II: 
Maximal lepton mixing 
($\sin^2\beta=0.5$). In both cases we set $\sin^2\phi=0.04$, since it 
corresponds to the minimal value of $x$.
Also, in Table 3 we tabulate the predictions of our model for 
various processes and for two cases. Case I: 
No lepton mixing ($\sin^2\beta=0.0$) and 
$x=20$. Case II: Maximal lepton mixing ($\sin^2\beta=0.5$) and $x=48$. 
For both cases we set $\sin^2\phi=0.04$.

Under the scenario of up-type quark mixing, 
there will be no non-standard effect present in the 
hadronic decays of $K$, $D$ and $B$ mesons.
This is because in the pure hadronic charged-current interaction, the
new physics effect is only expected in processes that involve the 
$b$ quark and where $\Delta B$ vanishes. 
Furthermore, the present data of semi-leptonic $b$-quark decays 
is not accurate enough to further constrain the model, though
it can be improved in the $B$ factories.
Under this scenario, the unitarity condition of the CKM matrix 
is modified, but its change is extremely small for the values of $x$ that
agree with $Z$-pole data.
In this case,
this model does not modify either the $B^0$-$\overline{B^0}$
or the $K^0$-$\overline{K^0}$ mixing at tree level. 
Although FCNC decay of charm meson is expected to be modified, 
the non-standard effect is very small because of the natural suppression
imposed by the tree level FCNC couplings (which are the product of 
CKM matrix elements).
With enough data in future experiments, the measurement of the partial 
decay widths of $\Upsilon(1S)$ into the $\tau^+ \tau^-$, $\mu^+ \mu^-$,
and $\mu^\pm \tau^\mp$ modes can further test the model.
Furthermore, it can also
modify the $K^0$-$\overline{K^0}$, $B^0$-$\overline{B^0}$ mixing
and the decay width of $b \rightarrow s \gamma$ at one-loop level.
However, the one-loop effects are small compared to the SM predictions and do 
not exceed the level of $0.4\%$ of the SM values.

Under the general mixing scenario,
the model requires two additional free (real) parameters, 
 although additional phases can be introduced to 
generate a new source of CP violation.
Depending on the values of the parameters, 
sizable effects in various FCNC processes are expected.
Therefore, low-energy data can also test the model with a general
mixing scenario. In Table 4 we summarize the results of the 
general mixing scenario by giving the constraints on the mixing 
parameters as extracted from different experiments. 
The general 
mixing scenario also allows lepton number violation processes, such as 
$B_{s,d}\rightarrow \mu^\pm \tau^\mp$ and 
$b \rightarrow s \mu^\pm \tau^\mp$.
Since their branching ratios are of the same order as those for the 
$\tau^+\tau^-$ mode, they can be very useful for further testing the model.

It is interesting to notice that in the SM
neither $L_u$ nor $L_d$ can be separately determined, and only
the CKM matrix $V$, which is the product of 
$L_u^\dagger$ and $L_d$, can be measured experimentally.
However, in this model, the elements in the third column of 
the $L_{u,d}$ mixing matrices can be determined, and can be 
further constrained by including other low-energy data.
Unfortunately, in general, those observables also
depend on some other
parameters, such as $\sin^2\phi$ and $\sin^2\beta$, of the model.

\section{Acknowledgments}
E.M. would like to thank K. Hagiwara, Y. Okada,
for useful discussion and comments. He also thanks KEK for 
the kind hospitality, where part of this work was done, 
and the Matsumae International Foundation for their
Fellowship to support his visit in Japan. 
This work  was supported in part
by the U.S.~NSF under grant PHY-9802564.

\newpage

\section*{Table Captions}
Table. 1.\\
Experimental data and predicted values of various electroweak
observables in the SM and the proposed model
(with different choices of parameters),
for $\alpha_s=0.118$, $m_t=175$ GeV and $m_H=100$ GeV.

\noindent
Table. 2.\\
The lower bound on $x$ derived from various decay processes for the 
  proposed model with the $d$-quark mixing scenario.  
  Case I: $\sin^2\beta=0$, $\sin^2\phi=0.04$. 
  Case II: $\sin^2\beta=0.5$, $\sin^2\phi=0.04$.

\noindent
Table. 3.\\
Predictions of various decay rates and mixing in the SM and the 
  proposed model with the $d$-quark mixing scenario.  
  Case I: $\sin^2\beta=0$, $x=20$, $\sin^2\phi=0.04$. 
  Case II: $\sin^2\beta=0.5$, $x=48$, $\sin^2\phi=0.04$.

\noindent
Table. 4.\\
Constraints on the quark mixing parameters from 
various decay processes for the 
 proposed model with the general mixing scenario.

\newpage

\section*{Figure Captions}

Fig. 1.\\
     {The lower bound on the heavy $Z^\prime$ mass as a function of 
     $\sin^2\phi$ at the $2\sigma$ level. 
     Solid curve: including all $Z$-pole data and assuming no lepton mixing.
     Dashed curve: including all $Z$-pole data and assuming maximal lepton 
     mixing ($\sin^2\beta=0.5$).
     Dotted curve: only including the hadronic measurements in the fit and 
     assuming no lepton mixing.}\\

\noindent
Fig. 2.\\
{The lower bound on the parameter $x$ as a function of 
 $\sin^2\phi$ at the $2\sigma$ level.\\
Solid curve: including all $Z$-pole data and assuming no lepton mixing.\\
Dashed curve: Including all data and assuming maximal lepton mixing 
($\sin^2\beta=0.5$).\\
Dotted curve: Including hadronic data only and assuming no lepton mixing.}\\

\newpage
\begin{table}
\caption{Experimental data and predicted values of various electroweak
observables in the SM and the proposed model
(with different choices of parameters),
for $\alpha_s=0.118$ with $m_t=175$ GeV and $m_H=100$ GeV.
Case  a: $\sin^2\beta=0$, $\sin^2\phi=0.04$, $x=20$, $M_Z^\prime=1.9$ TeV,
$\Gamma_Z^\prime=490$ GeV.
Case  b: $\sin^2\beta=0.5$, $\sin^2\phi=0.04$, $x=48$ 
(equivalently, $M_Z^\prime=2.8$ TeV, $\Gamma_Z^\prime=760$ GeV)
Case  c: $\sin^2\beta=0.0$, $\sin^2\phi=0.2$, $x=100$ 
(equivalently, $M_Z^\prime=2$ TeV, $\Gamma_Z^\prime=100$ GeV)}

\begin{tabular}{|c|c|c|c|c|c|}  \hline\hline
Observables&Experimental data&SM
&\multicolumn{3}{|c|} {\mbox {The model}}\\ \hline
 Included in fit  &          &  & a & b & c  \\ \hline
\underline{LEP1}  &                   
&             &    &  &  \\ 
$g_V(e)$          &   $-0.0367\pm 0.0015$   &-0.0374 
&-0.0374  & -0.0374  &-0.0375\\
$g_A(e)$          &   $-0.50123\pm 0.00044$ &-0.50142 
&-0.50140 & -0.50141 &-0.50132\\
$g_V(\mu)/g_V(e)$ &   $1.02\pm 0.12$        & 1.00   
&1.00     & 1.01     &  1.02   \\
$g_A(\mu)/g_A(e)$ &   $ 0.9993\pm 0.0017$   & 1.0000 
&1.0000   & 1.0004 &  1.0000 \\
$g_V(\tau)/g_V(e)$&   $0.998\pm 0.060$      & 1.000  
&1.027    & 1.006 & 1.027 \\
$g_A(\tau)/g_A(e)$&   $0.9996\pm 0.0018$    & 1.0000 
&1.0020   & 1.0004& 1.0020\\
$\Gamma_Z ({\rm{GeV}})$      & $2.4948\pm 0.0025$  & 2.4972 
& 2.4999  & 2.4983 & 2.4992\\  
$R_e$           & $20.757\pm 0.056 $  & 20.747 
& 20.770 & 20.757 & 20.770\\  
$R_\mu$         & $20.783\pm 0.037 $  & 20.747 
& 20.770 & 20.738 & 20.770\\
$R_\tau$        & $20.823\pm 0.050 $  & 20.795 
& 20.730 & 20.786 & 20.730 \\
$\sigma_h^0 (nb)$    & $41.486\pm 0.053 $  & 41.474 
& 41.422 & 41.452 & 41.422 \\
$A_e$           & $0.1399\pm  0.0073$  & 0.1484 
& 0.1485 & 0.1484 & 0.1487 \\
$A_\tau$        & $0.1411\pm 0.0064$   & 0.1484 
& 0.1521 & 0.1492  & 0.1523 \\
$A^{FB}_e$      & $0.0160\pm 0.0024$   & 0.0165 
& 0.0165 & 0.0165 & 0.0166\\
$A^{FB}_\mu$    & $0.0163\pm 0.0014$   & 0.0165 
& 0.0165 & 0.0166 & 0.0166\\
$A^{FB}_\tau$   & $0.0192\pm 0.0018$   & 0.0165 
& 0.0169 & 0.0166 & 0.0170\\
$R_b$           & $0.2170\pm 0.0009$   & 0.2157 
& 0.2165 & 0.2160 & 0.2165\\
$R_c$           & $0.1734\pm 0.0048$   & 0.1721 
& 0.1719 & 0.1720 & 0.1719\\
$A_{FB}^c$      & $0.0741\pm 0.0048$   &0.0744  
& 0.0744 & 0.0744 & 0.0746 \\ \hline
\underline{SLD}   & 
&             &   &  &         \\
$A_b$           & $0.900\pm 0.050$     &0.935   
& 0.935  & 0.935 & 0.935\\
$A_c$           & $0.650\pm 0.058$     &0.668   
& 0.668  & 0.668 & 0.668     \\ \hline
\underline{Tevatron + LEP2} &    
&          &        &   &    \\ 
$M_W ({\rm{GeV}})$     & $80.430\pm 0.084$  & 80.402 
& 80.403 & 80.403 &80.409 \\ \hline
\underline{Not included in fit} &  &  & & & \\
$A_{FB}^b$      & $0.0984\pm 0.0024$   &0.1040  
& 0.1041 & 0.1040 & 0.1043 \\
$A_{LR}$        & $0.1547\pm 0.0032$   & 0.1484 
& 0.1485 & 0.1484 &0.1487\\ \hline
\end{tabular}
\end{table}
\noindent

\newpage
\begin{table}
\caption{
The lower bound on $x$ derived from various decay processes for the 
  proposed model with the $d$-quark mixing scenario.  
  Case I: $\sin^2\beta=0$, $\sin^2\phi=0.04$. 
  Case II: $\sin^2\beta=0.5$, $\sin^2\phi=0.04$.}

\vspace{0.1in}

\begin{tabular}{|c|c|c|}  \hline\hline
Process  &
\multicolumn{2}{|c|} {\mbox{$x >$}} \\ \hline
    --    &  I  & II \\ \hline
$Z$-Pole data & 20  & 20 \\
${\rm{Br}}(\tau^-\rightarrow \mu^- \overline{\nu_\mu} \nu_\tau)/
{\rm{Br}}(\tau^-\rightarrow e^- \overline{\nu_e} \nu_\tau)$
& 0 & 48 \\
${\rm{Br}}(\tau^-\rightarrow\mu^- \mu^+ \mu^-)$  
& 0   & 37 \\
${\rm{Br}}(\tau\rightarrow \mu \gamma)$  
& 0 & 3 \\
${\rm{Br}}(K^+ \rightarrow \pi^+ \nu \overline{\nu})$  
& 7 & 7 \\
${\rm{Br}}(b \rightarrow s \mu^+\mu^-)$ 
&0 & 19 \\
${\rm{Br}}(b \rightarrow s \nu\overline{\nu})$ 
& 15 & 15 \\
${\rm{Br}}(B_d \rightarrow \mu^+\mu^-)$ 
&0 & 1 \\
${\rm{Br}}(B_s \rightarrow  \mu^+\mu^-)$ 
& 0 &  4 \\
\hline
\end{tabular}
\end{table}

\newpage
\begin{table}
\caption{Predictions of various decay rates and mixing in the SM and the 
  proposed model with the $d$-quark mixing scenario.  
  Case I: $\sin^2\beta=0$, $x=20$, $\sin^2\phi=0.04$. 
  Case II: $\sin^2\beta=0.5$, $x=48$, $\sin^2\phi=0.04$.}

\vspace{0.1in}

\begin{tabular}{|c|c|c|c|c|}  \hline\hline
Process &  Data & SM  &
\multicolumn{2}{|c|} {\mbox{d-type mixing}} \\ \hline
    --    &  --    & --  &  I  & II \\ \hline
$\frac{{\rm{Br}}(\tau^-\rightarrow \mu^- \overline{\nu_\mu} \nu_\tau)}
{{\rm{Br}}(\tau^-\rightarrow e^- \overline{\nu_e} \nu_\tau)}$
& $0.976\pm 0.006$       & 0.9729 
& 0.9729 & 0.9881 \\
${\rm{Br}}(\tau^-\rightarrow \mu^- \mu^+ \mu^-)$  
&$ < 1.9 \times 10^{-6}$   & 0   
& 0   &  $1.1\times 10^{-6}$ \\
${\rm{Br}}(\tau \rightarrow \mu \gamma)$
& $<4.2\times 10^{-6}$ & 0 
& 0  & $ 1.7\times 10^{-8}$\\
${\rm{Br}}(K^0_L \rightarrow \mu^+ \mu^-)$  
& $(7.2\pm 0.5)\times 10^{-9}$ & $\sim 7\times 10^{-9}$
& $1.3\times 10^{-10}$ & $3.4\times 10^{-9}$\\   
${\rm{Br}}(K^+ \rightarrow \pi^+ \nu \overline{\nu})$  
& $4.2 ^{+9.7}_{-3.5}\times 10^{-10}$ & $(9.1 \pm 3.8)\times 10^{-11}$ &
$2.8 \times 10^{-10}$ & $4.8 \times 10^{-11}$ \\
$\Delta M_K \, (ns^{-1})$ 
&$5.311\pm 0.019$ & $2.23 \sim 7.43$ & 
$2.6\sim 8.9$ & $2.4 \sim 8.0$\\
%
%
$\Delta M_{B_s} \, (ps^{-1})$       
& $>10.2 $ & $ 1\sim 15$  &  
$5 \sim 69$  & $3 \sim 37$ \\
${\rm{Br}}(b \rightarrow s \mu^+\mu^-)$ 
& $<5.8 \times 10^{-5}$ & $\sim 7\times 10^{-6}$ & 
$1.6\times 10^{-7}$ & $9.2 \times 10^{-6}$ \\
${\rm{Br}}(b \rightarrow s \nu\overline{\nu})$ 
& $<3.9 \times 10^{-4}$ & $\sim 4.2\times 10^{-5}$ & 
$2.3\times 10^{-4}$ & $4.0 \times 10^{-5}$ \\
${\rm{Br}}(b \rightarrow s \mu^\pm \tau^\mp)$ 
& ? &  0 
& 0 & $2.0 \times 10^{-5}$ \\
${\rm{Br}}(B_d \rightarrow \mu^+\mu^-)$ 
& $<8.6 \times 10^{-7}$ &  $ 2.1\times 10^{-10}$ & 
$3.2\times 10^{-11}$ & $8.8 \times 10^{-10}$ \\
${\rm{Br}}(B_s \rightarrow  \mu^+\mu^-)$ 
& $<2.6 \times 10^{-6}$ & $ 4.3 \times 10^{-9}$ & 
$6.1\times 10^{-10}$  & $1.7 \times 10^{-8}$ \\
${\rm{Br}}(B_d \rightarrow  \mu^\pm \tau^\mp)$ 
& $<8.3 \times 10^{-4}$ & 0 & 
0  & $4.0 \times 10^{-7}$ \\
${\rm{Br}}(B_s \rightarrow  \mu^\pm \tau^\mp)$ 
& ? & 0 & 
0  & $7.7 \times 10^{-6}$ \\
${\rm{Br}}(B_d \rightarrow \tau^+\tau^-)$ 
& ? &  $ 4.3\times 10^{-8}$ & 
$2.6\times 10^{-6}$ & $1.0 \times 10^{-7}$ \\
${\rm{Br}}(B_s \rightarrow  \tau^+\tau^-)$ 
& ? & $ 9.1 \times 10^{-7}$ & 
$5.0\times 10^{-5}$  & $2.0 \times 10^{-6}$ \\
$\Upsilon(1S) \to \mu^\pm \tau^\mp $ 
&? & 0 &
0 & $ 4\times 10^{-10}$ \\       
\hline
\end{tabular}
\end{table}

\newpage
\begin{table}
\caption{Constraints on the quark mixing parameters from various 
decay processes for the 
  proposed model with the general mixing scenario.}

\vspace{0.1in}

\begin{tabular}{|c|c|c|c|}  \hline\hline
Process   &   $\sin^2\beta$  & $\sin^2\phi$ & Constraint  \\ \hline
${\rm{Br}}(K^+\rightarrow \pi^+ \nu\overline{\nu})$ 
& independent & independent & 
$|d_{31}d_{32}|/x \lsim 1.0\times 10^{-4}$ \\
${\rm{Br}}(b\rightarrow s \nu\overline{\nu})$ 
& independent & independent &
$|d_{32}d_{33}|/{x}\lsim 2.9\times 10^{-3}$ \\
${\rm{Br}}(b\rightarrow s \mu^+\mu^-)$ 
& 0.5 & 0.04 &
$|d_{32}d_{33}|/{x}\lsim 2.3\times 10^{-3}$ \\
${\rm{Br}}(B_d\rightarrow \mu^+\mu^-)$ 
& 0.5& 0.04 &
$|d_{31}d_{33}|/x \lsim 9.1 \times 10^{-3}$ \\
${\rm{Br}}(B_s\rightarrow \mu^+\mu^-)$ 
& 0.5 & 0.04 &
$|d_{32}d_{33}|/{x}\lsim 1.2 \times 10^{-2}$ \\ \hline
\end{tabular}
\end{table}


\newpage

\begin{thebibliography}{99}

\bibitem{hill1}  C. T. Hill, Phys. Lett. {\bf B266}, 419 (1991);
                             Phys. Lett. {\bf B345}, 483 (1995);
S.P. Martin, Phys. Rev. {\bf D46}, 2197 (1992); 
             Phys. Rev. {\bf D45}, 4283 (1992);
             Nucl. Phys. {\bf B398}, 359 (1993);
M. Lindner and D. Ross, Nucl. Phys. {\bf B370}, 30 (1992);
R. Bonisch, Phys. Lett. {\bf B268}, 394 (1991);
C.T. Hill, D. kennedy, T. Onogi, and H.L. Yu, Phys. Rev. {\bf D47}, 
2940 (1993).

\bibitem{hill2} G. Buchalla, G. Burdman, C.T. Hill, and D. Kominis,
  Phys. Rev. {\bf D53}, 5185 (1996).

\bibitem{chiv} R.S. Chivukula, E.H. Simmons, J. Terning,
 Phys. Lett. {\bf B331}, 
383 (1994); ibid, Phys. Rev. {\bf D53}, 5258 (1996). 

\bibitem{lima} 
X. Li and E. Ma, Phys. Rev. Lett. {\bf 47},  
1788 (1981); E. Ma,  X. Li, and S.F. Tuan 
Phys. Rev. Lett. {\bf 60}, 
495 (1988); X. Li and E. Ma, Phys. Rev. {\bf D46}, 1905 (1992);  
J. Phys. {\bf G19}, 1265 (1993). 

\bibitem{ehab5}  E. Malkawi, T. Tait, and C.--P. Yuan. Phys. Lett. {\bf B385}, 
304 (1996).

\bibitem{muller}  D.J. Muller and S. Nandi, Phys. Lett. {\bf B383}, 345 (1996).

\bibitem{lee}  J.C. Lee and K.Y. Lee, Phys. Rev. {\bf D58}, 115001 (1998);
  J.C. Lee and K.Y. Lee, and J.K. Kim,  Phys. Lett. {\bf B424}, 133 (1998).

\bibitem{andrea}  A. Donini, F. Feruglio, J. Matias, F. Zwirner, Nucl. Phys. 
{\bf B507}, 51 (1997).

\bibitem{lep} The LEP Electroweak Working group and the 
SLD Heavy Flavor Group,
  CERN-PPE/97-154, December 1997. 

\bibitem{data}  Review of Particle Properties, {\it The European Physical 
Journal } {\bf C3}, 1 (1998), and the www page, http://pdg.lbl.gov. 

\bibitem{hagiwara} K. Hagiwara, Annu. Rev. Nucl. Part. Sci. {\bf 48},
463 (1998).

\bibitem{ununi}  H. Georgi, E.E. Jenkins, and E.H. Simmons, Phys. Rev. Lett 
{\bf 62}, 2789 (1989); {\em ibid.}, Nucl. Phys. {\bf B331}, 541 (1990);\\ 
R.S. Chivukula, E.H. Simmons and J. Terning, Phys.Lett. {\bf B346}, 284
(1995).  

\bibitem{park}I. Park, Nucl. Phys. Proc. Supp. {\bf 65}, 136 (1998);
M. Schmidtler, Nucl. Phys. Proc. Supp. {\bf 65}, 142 (1998).

\bibitem{bnl_amu} B.L. Roberts et al, The E821 Collaboration. Published in the 
proceedings of the 28th International Conference on High-Energy Physics (ICHEP 96), 
Warsaw, Poland. World Scientific, River Edge, NJ, 1997, pp. 1035. 
R.M. Carey, Phys. Rev. Lett. {\bf 82}, 1632 (1999).

\bibitem{amu_sm} Andrzej Czarnecki and Bernd Krause, 
Nucl. Phys. Proc. Suppl. {\bf 51C}, 148 (1996); and the references therein.

\bibitem{lashin} V. Antonelli, S. Bertolini, M. Fabbrichesi, and E.I. Lashin, 
Nucl. Phys. {\bf B493}, 281 (1997).

\bibitem{inami} T. Inami and C.S. Lim, Prog. Theor. Phys. 
{\bf 65}, 297 (1981).

\bibitem{ali1} A. Ali and B. Kayser, hep-ph/9806230.

\bibitem{ellis} J. Ellis and J.S. Hagelin, Nucl. Phys. {\bf B217}, 
189 (1983);
D. Rein and L.M. Sehgal, Phys. Rev. {\bf D39}, 3325 (1989);
J.S. Hagelin and L.S. Littenberg, Prog. Part. Nucl. Phys. {\bf 23}, 
1 (1989);
C.Q. Gang, I.J. Hsu, and Y.C. Lin, Phys. Lett. {\bf B355}, 569 (1995);
S. Fajfer, Nuov. Cim. {\bf A110}, 397 (1997).

\bibitem{e787} E787 Collaboration, Phys. Rev. Lett. {\bf 79}, 2204 (1997).

\bibitem{buras} A.J. Buras and R. Fleischer, hep-ph/9704376;
 A.J. Buras, hep-ph/9711217; 
 G. Buchalla and A.J. Buras, Nucl. Phys. {\bf B400}, 
  225 (1993); A. Ali, C. Greub and T. Mannel, DESY-93-016;
 G. Buchalla and A.J. Buras, M.E. Lautenbacher, 
Rev. Mod. Phys. {\bf 68}, 1125 (1996).  

 
\bibitem{ahmad} Mohammad R. Ahmady, Phys. Rev. {\bf D53}, 2843 (1996);
       

\bibitem{cleo1} CLEO Collaboration, Phys. Rev. Lett. {\bf 80}, 1150 (1998).

\bibitem{cleo2} CLEO Collaboration, Phys. Rev. Lett. {\bf 80}, 2289 (1998).

\bibitem{nardi} Y. Grossman, Z. Ligeti, E. Nardi,
Nucl. Phys. {\bf B465}, 369 (1996).

\bibitem{ali2} A. Ali, DESY 96-106, hep-ph/9606324, and the 
references therein.

\bibitem{abe} F. Abe et al. The CDF Collaboration, 
Phys. Rev. {\bf D57}, 3811 (1998). 

\bibitem{fuji} K. Fujikawa and A. Yamada, Phys. Rev. {\bf D49}, 5890 (1994). 

\end{thebibliography}
\end{document}